\documentclass[prb,twocolumn,floatfix,superscriptaddress,longbibliography]{revtex4-1}
\usepackage[pdftex,plainpages=false,colorlinks=true,linkcolor=blue, citecolor=blue, urlcolor=blue]{hyperref}
\usepackage{amsfonts}
\usepackage{amsmath}
\usepackage{amssymb}
\usepackage{graphicx}
\usepackage{natbib}
\usepackage{color}
\usepackage{placeins}
\usepackage{physics}

\begin{document}
 
\title{Thermodynamic and information-theoretic description of the Mott transition in the two-dimensional Hubbard model}

\author{C. Walsh}
\affiliation{Department of Physics, Royal Holloway, University of London, Egham, Surrey, UK, TW20 0EX}
\author{P. S\'emon}
\affiliation{Computational Science Initiative, Brookhaven National Laboratory, Upton, NY 11973-5000, USA}
\author{D. Poulin}
\affiliation{D\'epartement de physique and Institut quantique, Universit\'e de Sherbrooke, Sherbrooke, Qu\'ebec, Canada J1K 2R1}
\affiliation{Canadian Institute for Advanced Research, Toronto, Ontario, Canada, M5G 1Z8}
\author{G. Sordi}
\email[corresponding author: ]{giovanni.sordi@rhul.ac.uk}
\affiliation{Department of Physics, Royal Holloway, University of London, Egham, Surrey, UK, TW20 0EX}
\author{A.-M. S. Tremblay}
\affiliation{D\'epartement de physique and Institut quantique, Universit\'e de Sherbrooke, Sherbrooke, Qu\'ebec, Canada J1K 2R1}
\affiliation{Canadian Institute for Advanced Research, Toronto, Ontario, Canada, M5G 1Z8}

\date{\today}

\begin{abstract}
At the Mott transition, electron-electron interaction changes a metal, in which electrons are itinerant, to an insulator, in which electrons are localized. This phenomenon is central to quantum materials. Here we contribute to its understanding by studying the two-dimensional Hubbard model at finite temperature with plaquette cellular dynamical mean-field theory. 
We provide an exhaustive thermodynamic description of the correlation-driven Mott transition of the half-filled model by calculating pressure, charge compressibility, entropy, kinetic energy, potential energy and free energy across the first-order Mott transition and its high-temperature crossover (Widom line). The entropy is extracted from the Gibbs-Duhem relation and shows complex behavior near the transition, marked by discontinuous jumps at the first-order boundary, singular behavior at the Mott endpoint and inflections marking sharp variations in the supercritical region. The free energy allows us to identify the thermodynamic phase boundary, to discuss phases stability and metastability, and to touch upon nucleation and spinodal decomposition mechanisms for the transition. 
We complement this thermodynamic description of the Mott transition by an information-theoretic description. We achieve this by calculating the local entropy, which is a measure of entanglement, and the single-site total mutual information, which quantifies quantum and classical correlations. These information-theoretic measures exhibit characteristic behaviors that allow us to identify the first-order coexistence regions, the Mott critical endpoint and the crossovers along the Widom line in the supercritical region. 
\end{abstract}

\date{\today}

\maketitle

\section{Introduction}
At the Mott metal-insulator transition, the Coulomb interaction in a half-filled band competes with the kinetic energy to change the collective behavior of the electrons from itinerant to localised~\cite{mott,ift}. As unconventional superconductivity and other exotic quantum states occur in proximity to Mott insulators, describing the Mott transition in quantum materials remains a central programme in condensed matter physics~\cite{keimer:Rev2017, basov:Rev2017, ift, anderson:1987}, one that catalizes key advances in experimental and theoretical techniques alike. On the experimental side, new approaches to study correlated systems emerged, such as ultracold atoms in optical lattices~\cite{Jordens:2008, Schneider:2008, bdzRMP, Hofrichter:PRX2016, Cocchi:PRL2016}, and more recently twisted two-dimensional superlattices~\cite{Cao-Graphene1:2018, Cao-Graphene2:2018}. On the theoretical side, the Hubbard model is the simplest model that captures the Mott transition. Nevertheless, understanding the Mott transition even within simple models is a difficult task because it is a non-perturbative phenomenon, thereby preventing the use of known analytical methods. Dynamical mean-field theory~\cite{rmp} and its extensions~\cite{maier, kotliarRMP, tremblayR} emerge as powerful tools that provide a non-perturbative approach to the Mott transition. Theoretical progress in the description of the Mott transition in the Hubbard model in two dimensions, where local quantum fluctuations play together with short-range spatial correlations, is particularly challenging~\cite{LeBlancPRX}. 

While key results have been obtained, here we take advantage of algorithmic improvement and extensive computer resources to provide a detailed thermodynamic description of the Mott transition in the half-filled two-dimensional Hubbard model within cellular dynamical mean-field theory. We reveal the landscape of pressure, charge compressibility, thermodynamic entropy, kinetic energy, potential energy and free energy across the Mott transition and its high-temperature crossover. Key thermodynamic signatures in these observables, such as inflection points in the entropy, are sometimes visible at high temperature only when an unprecedented level of accuracy is attained.  

Knowledge of the thermodynamic entropy allows us to complement the thermodynamic description of the Mott transition by a description based on quantum information theory (see also our companion letter~\cite{walshSl}). Although Jaynes~\cite{Jaynes:1957, Jaynes:1957b} already unveiled the links between information theory and thermodynamics, it is only in the last decade that a classification of phase transitions in correlated many-body systems using information-theoretic tools become a major research direction~\cite{amicoRMP2008, EisertRMP, VedralRMPDiscord}. Even more recently, ultracold atom experiments have become able to access information-theoretic measures of quantum correlations~\cite{greinerNat2015, Cocchi:PRX2017}, opening the avenue to quantify and manipulate quantum correlations in many-body quantum systems, and calling for further theoretical investigations on Hubbard-type models. Here we expand on our companion letter~\cite{walshSl} by providing further results on the relation between Mott transition and two measures of quantum correlations, the local entropy and total mutual information. The former is a measure of entanglement, whereas the latter, defined as the difference between local entropy and thermodynamic entropy, quantifies all classical and quantum correlations. We reveal that these measures are able to detect the first-order character of the transition, the critical behavior near the Mott endpoint and the supercritical crossover emerging from the endpoint. 

We describe the model and method in Sec.~\ref{sec:ModelAndMethod}. Section~\ref{sec:Docc} presents the phase diagram mainly through the behavior of double occupancy. It also shows the occupation as a function of chemical potential, key to the calculation of the entropy. The charge compressibility can also be extracted from it. The Gibbs-Duhem relation is used in Sec.~\ref{sec:GibbsDuhem} to find pressure and entropy. It also gives us the opportunity to do an accuracy check by providing an alternate way to obtain the kinetic energy. Questions of thermodynamic stability are explored in Sec.~\ref{sec:Stability}. In Sec.~\ref{sec:Entanglement} we characterize the Mott transition using two information theoretic measures, the local entropy and the mutual information. Sec.~\ref{sec:Conclusions} summarizes our findings. Appendix~\ref{sec:CriticalS} recalls the scaling behavior of the entropy at the Mott critical endpoint. 

\section{Methodology}
\label{sec:ModelAndMethod}

We study the single-band Hubbard model on the square lattice in two dimensions:
\begin{equation}
H=-\sum_{ij\sigma}t_{ij}c_{i\sigma}^\dagger c_{j\sigma}
+U\sum_{i} n_{i\uparrow } n_{i\downarrow }
-\mu\sum_{i\sigma} n_{i\sigma}, 
\label{eq:HM}
\end{equation}
where $c^{\dagger}_{i\sigma}$ and $c_{i\sigma}$ operators create and annihilate an electron of spin $\sigma$ on site $i$, $n_{i\sigma}=c^{\dagger}_{i\sigma}c_{i\sigma}$ is the number operator, $U$ is the onsite Coulomb repulsion, and $\mu$ is the chemical potential. We take hopping amplitudes $t_{ij}$ between nearest neighbors only and set  $t_{ij}=t=1$ as our energy unit. 

One of the most advanced methods for a theoretical treatment of this model is cellular dynamical mean-field theory (CDMFT)~\cite{maier, kotliarRMP, tremblayR}, which is a cluster extension of DMFT~\cite{rmp}. 
This theory provides a framework for understanding local quantum fluctuations generated by the interaction $U$ on the same footing as the short-range spatial correlations. CDMFT does so by taking a cluster of lattice sites, here a $2\times2$ plaquette, out of the lattice and by replacing the missing lattice environment by a self-consistent bath of noninteracting electrons. 

To solve the impurity (cluster in a bath) problem, we use continuous-time quantum Monte Carlo method (CTQMC)~\cite{millisRMP}, based on expansion of the hybridization between cluster and bath (CT-HYB). The Lazy-skip list algorithm~\cite{patrickSkipList} is implemented for speed. Self-consistency is attained using an iterative procedure. Convergence is reached typically within 50 iterations, but hundreds are necessary close to phase boundaries. Once convergence is attained, we take averages over at least the last 30 CDMFT iterations and the resulting root mean square deviation on local quantities, such as the occupation $n$ and the double occupation $D$, is on the fifth digit. The number of Monte Carlo updates during each iteration is of order $10^9$.

\section{Phase diagram}
\label{sec:Docc}

\begin{figure*}
\centering{
\includegraphics[width=1.0\linewidth]{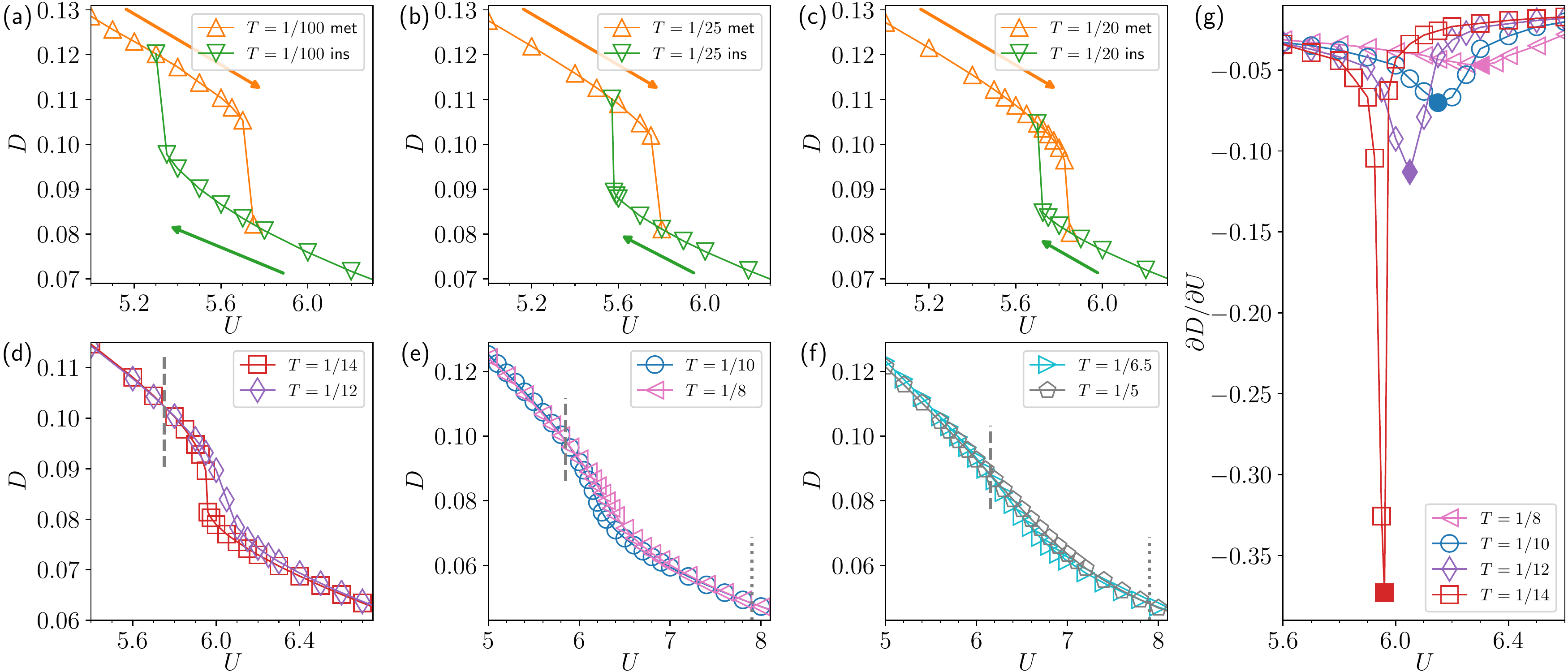}
}
\caption{(a)-(f) Isothermal double occupancy $D$ versus $U$ for several temperatures, at $n=1$. Upper panels show $T<T_c$, where $D(U)|_T$ displays hysteretic behavior. Hysteresis loops are obtained by sweeping the interaction strength $U$ from left to right and from right to left. Arrows indicate the sweep direction. The jumps in the double occupancy mark the spinodal points. Lower panels show $D(U)$ for $T>T_{\rm c}$.  In the temperature range considered in panels (d)-(f), each $D(U)|_T$ has an inflection point, where the concavity changes from negative to positive. At such inflection points the slope of the curve becomes steeper upon decreasing $T$ towards $T_{\rm c}$. This behavior is quantified in panel (g), where $(\partial D / \partial U)_T$ is plotted versus $U$ for several temperatures above $T_{\rm c}$. The first derivative $(\partial D / \partial U)_T$ shows a minimum that sharpens and whose value becomes more pronounced with decreasing $T$. The locus of these minima defines the Widom line $T_W$ in the $T-U$ phase diagram of Fig.~\ref{figS-PhaseDiagram}. In panels (d),(e), (f), the grey dashed and dotted lines indicates the crossing of the isotherms, i.e. where $(\partial D/ \partial T)_U =0$ (see Figure~\ref{figS-s2} and discussion therein). 
}
\label{figS-docc}
\end{figure*}
\begin{figure}
\centering{
\includegraphics[width=1\linewidth]{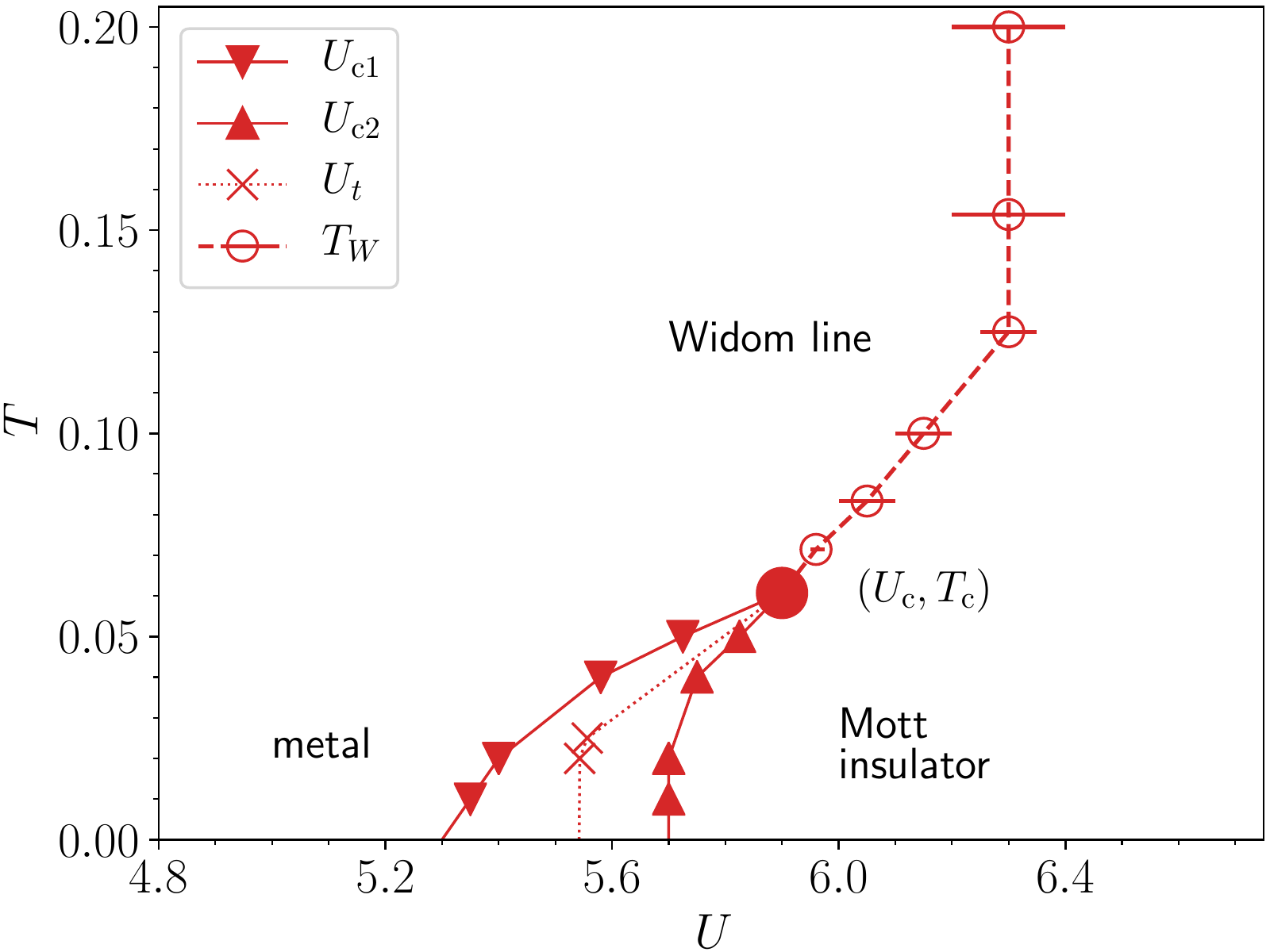}
}
\caption{Temperature-interaction strength phase diagram of the half-filled 2D Hubbard model within plaquette CDFMT. Lines with full triangles mark the spinodal lines $U_{c1}$ and $U_{c2}$, where the insulating and metallic solutions cease to exist, respectively. They are determined by the position of the jumps in $D(U)|_T$. Dotted line with crosses indicates the thermodynamic transition curve, $U_t$, obtained by the crossing of the grand potential (see Figure~\ref{figS-F} and discussion therein). Full circle indicates the critical Mott endpoint, where the coexistence disappears. Dashed line with open circles marks the Widom line $T_W$, i.e. the supercritical crossover determined by the locus of inflections of the double occupancy versus $U$ [i.e.  min$(\partial D /\partial U)_T$]. All lines are guides to the eye. 
}
\label{figS-PhaseDiagram}
\end{figure}

Let us consider the two-dimensional Hubbard model at half filling ($n=1$). The phase diagram at half-filling is determined by temperature $T$ and interaction strength $U$. Numerical calculations based on cluster extensions of DMFT have unveiled a simple yet rich phase diagram in the $T-U$ plane: in the normal state at low temperature and intermediate interaction the system undergoes a first-order transition between a metal and a Mott insulator. This first-order transition ends in a critical endpoint at $(U_{\rm c}, T_{\rm c})$, where the transition becomes continuous. The supercritical region, i.e. the region in the $T-U$ plane at temperature higher than the endpoint, displays interesting crossovers, understood through the concept of the Widom line. The Widom line is a crossover line defined as the locus of the maxima of the correlation length emanating from the endpoint into the supercritical region~\cite{water1, supercritical, ssht}. Indeed, at the critical endpoint the correlation length diverges, and asymptotically close to the endpoint all response functions are proportional to powers of the correlation length, so the extrema of the response functions converge asymptotically close the endpoint~\cite{water1, supercritical}. This concept was formulated in the context of fluids~\cite{water1, supercritical} and extended to electronic fluids in Ref.~\onlinecite{ssht}. 

In this section we revisit the $T-U$ phase diagram at $n=1$ with state of the art plaquette CDMFT calculations by focusing on the behavior of the double occupancy and single occupancy. The purpose of this section is twofold. First, the results obtained form the starting point of our discussion of thermodynamic quantities in Sections~\ref{sec:GibbsDuhem} and \ref{sec:Stability} that help throw new light into the nature of the Mott transition. Second, our analysis improves the determination of phase boundaries, endpoint and Widom line with a level of accuracy of about one percent.

\subsection{Double occupancy}

First, we construct the phase diagram of the two-dimensional Hubbard model in the $T-U$ plane at $n=1$. We focus on three key aspects: the first-order nature of the Mott transition, the critical Mott endpoint and the Widom line. We achieve these objectives by carefully computing the isothermal double occupancy $D$ as a function of interaction strength $U$ for different values of temperature in the range $1/100  \le T \le 1/5$, where the Mott transition and its associated crossovers lie (Figure~\ref{figS-docc}). Figure~\ref{figS-PhaseDiagram} shows the resulting phase diagram in the $T-U$ plane. We calculated about 500 points in the $T-U$ plane.

\subsubsection{First-order transition}
For $T<T_{\rm c}$, on sees in Figs~\ref{figS-docc}(a)-(c) that $D(U)|_T$ shows hysteresis loops, which are the hallmark of the first-order nature of the Mott transition. As expected, the Mott insulator has less double occupancies than the metal. Hysteresis has been obtained by sweeping up in $U$ (orange up triangles) and sweeping down in $U$ (green down triangles). The discontinuous jumps in $D(U)|_T$ signal the disappearance of the insulating state at $U_{\rm c1}$ and of the metallic state at $U_{\rm c2}$. Thus, by performing $U$ sweeps at different temperatures, we can obtain the spinodal lines $U_{\rm c1}(T)$ and $U_{\rm c2}(T)$ [lines with down and up triangles, respectively, in Fig.~\ref{figS-PhaseDiagram}(a)]. Hysteresis loops vary with $T$, decreasing in size with increasing $T$. Therefore the coexistence region in Fig.~\ref{figS-PhaseDiagram}(a) shrinks with increasing $T$. 

\subsubsection{Widom line}
For $T>T_{\rm c}$ (the so-called supercritical region), the isotherms $D(U)|_T$ are single-valued, and monotonically decreasing functions of $U$, as can be seen in Figs.~\ref{figS-docc}(d)-(f). Far away from $T_{\rm c}$, at $T \approx 0.2$ or $\approx 3.33 T_{\rm c}$, the isotherms $D(U)|_T$ start to develop an inflection point, where their curvature change from negative to positive. Moreover, $D(U)|_T$ at the inflection point becomes progressively steeper with decreasing $T$ towards $T_{\rm c}$. As a results, $(\partial D / \partial U)_T$ develops a minimum, which sharpens and whose value increases, i.e. becomes more negative, with progressively decreasing $T$ [Fig.~\ref{figS-docc}(g)], and eventually diverges at $T_{\rm c}$. 
The line connecting the values of $U$ corresponding to each of the inflection points in $D(U)|_T$ -as determined by the minima in Fig.~\ref{figS-docc}(g) - is our estimate for the Widom line in the $T-U$ plane, as indicated by circles in Fig.~\ref{figS-PhaseDiagram}. From the phase diagram it is then clear that the Widom line is a crossover line emanating out of the Mott endpoint into the supercritical region. 

To understand the significance of the Widom line, let us contrast the behavior of $D(U)$ below and above $T_{\rm c}$: in the same way that the divergence in $\partial D / \partial U$ at the Mott endpoint is the {\it precursor} of the phase coexistence below $T_{\rm c}$, so the line of inflections of $D(U)$ above $T_{\rm c}$ is the {\it precursor} of the Mott endpoint at $T_{\rm c}$. From high to low temperature therefore one has the sequence: the crossover line of minima in $(\partial D / \partial U)_T$ develops into a critical point at $(U_{\rm c}, T_{\rm c})$, which is then followed by a first-order transition at low temperature. Now, it is often the case that the Mott endpoint and the underlying first-order transition are masked by some broken symmetry phases, such as long-range antiferromagnetism~\cite{rmp, LorenzoAF} or superconductivity~\cite{sshtSC}. Because the Widom line emanates out of the critical endpoint and persists up to high temperature in the normal phase, the Widom line can be used {\it to extrapolate} the existence of the endpoint and its location. The possibility of gaining information on a hidden critical endpoint by using the supercritical crossover emanating from it is one of the key motivations behind the introduction of the concept of Widom line: in fluids, it was originally discussed that the Widom line might point to the existence of a liquid-liquid transition in supercooled water~\cite{water1}; in electronic fluids, we introduced it~\cite{ssht} to pinpoint the existence of a metal-metal transition beneath the superconducting dome in the doped 2D Hubbard model~\cite{ssht, sshtSC, sshtRHO}.

\subsubsection{Mott endpoint}

\begin{figure}
\centering{
\includegraphics[width=1\linewidth]{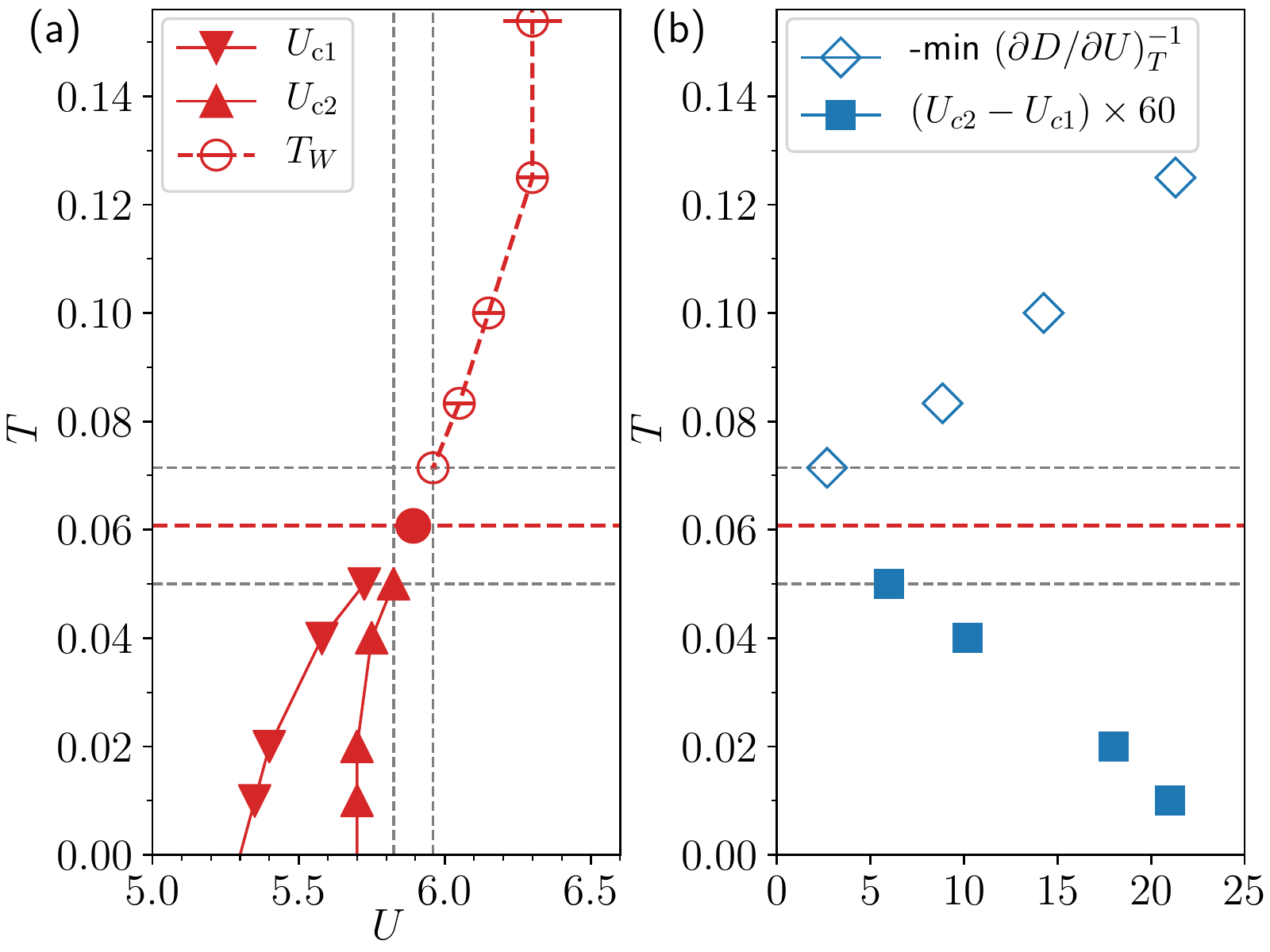}
}
\caption{(a) $T-U$ phase diagram. Lines with triangles: $U_{c1}(T)$ (down triangles) and $U_{c2}(T)$ (up triangles). Dashed line with open red circles is the Widom line, $T_W$. Dashed grey lines places bounds for the location of the critical endpoint (red filled circle). Bounds on $T_c$ in the temperature range near the critical point (on the vertical scale) are shown in panel (b). We plot on the horizontal axis the size of the coexistence region $(U_{c2}-U_{c1})$ (times 60, for better visualisation)  (blue squares). At $T_c$ the coexistence disappears, so this value extrapolates to 0 at $T_c$. In addition, values of the -min$(\partial D/\partial U)^{-1}$ are plotted on the horizontal axis (blue diamonds). At the endpoint, $\partial D/ \partial U$ diverges, so $(\partial D/\partial U)^{-1}$ goes to zero. 
}
\label{figS-Tc}
\end{figure}
\begin{figure*}
\centering{
\includegraphics[width=1\linewidth]{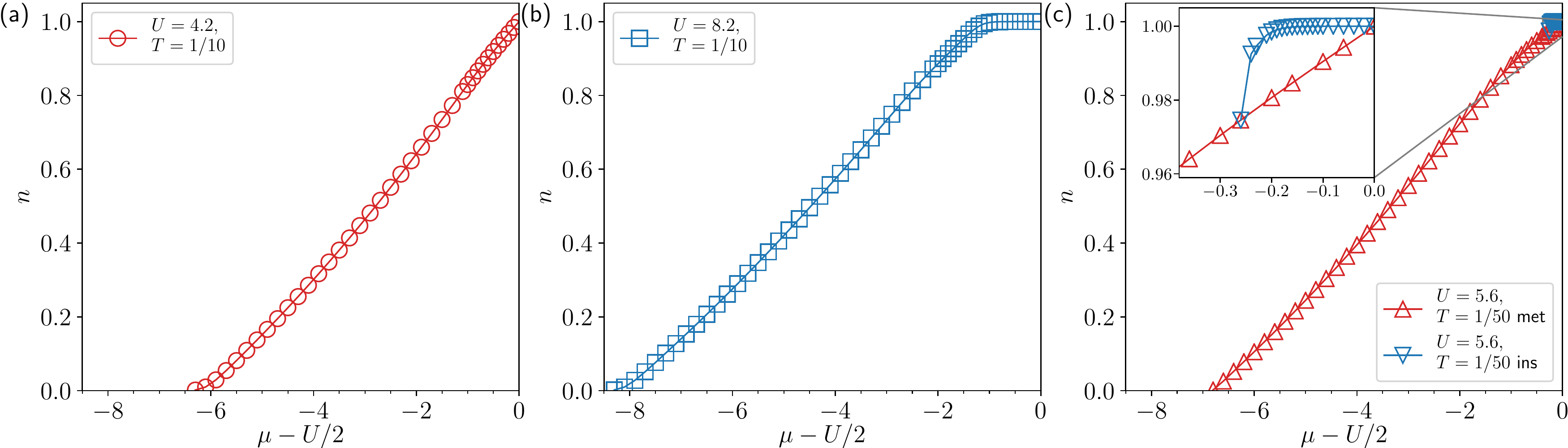}
}
\caption{Occupation $n(\mu)$ for (a) $U=4.2<U_{\rm c}$ and $T=1/10>T_c$, (b) $U=8.2>U_{\rm c}$ and $T=1/10>T_c$, (c) $n(\mu)$ for $U=5.6$ for $T=1/50<T_{\rm c}$, which lies in the coexistence region of the $T-U$ phase diagram: a metallic solution (up triangles) and an insulating solution (down triangles) coexist. On the $x$-axes we used the shifted chemical potential $\tilde{\mu}=\mu-U/2$.
}
\label{figS-n}
\end{figure*}

At $T=T_{\rm c}$, the isotherm $D(U)|_{T_{\rm c}}$ is continuous with an inflection point with vertical tangent at $U_{\rm c}$, resulting in a divergence in $\partial D / \partial U$. A detailed study~\cite{patrickCritical} has shown that at $T_{\rm c}$, $D(U)$ as a function of $U$ has critical behavior and scales as $-\text{sgn}(U-U_c)|(U-U_c)|^{1/\delta}$, with $\delta=3$ in CDMFT, and therefore $(\partial D / \partial U)_{T_{\rm c}}$ scales as $-|(U-U_c)|^{(1/\delta)-1}$ near $(U_{\rm c},T_{\rm c})$.  

The singular behavior in $D(U)$ at the Mott endpoint extends in the supercritical region in the form of inflection points, i.e. the Widom line: the divergence in $\partial D / \partial U$ at $T_c$ is replaced, for $T>T_c$, by a sharp minimum, which smears out and whose value  progressively moves away from $T_c$. Therefore, for $T>T_{\rm c}$, from the Mott endpoint it emerges a sharp crossover line, the Widom line. On the other hand, for $T<T_{\rm c}$, the Mott endpoint is the terminus of the finite-temperature first-order Mott transition, where the metallic and insulating phases merge into a single phase. 

These arguments lead to a natural way to estimate the Mott critical endpoint $(U_{\rm c},T_{\rm c})$. To find an upper bound for $T_{\rm c}$, we proceed as follows. By construction, the Widom line is made of the minima of $\partial D/ \partial U$, whose magnitude becomes more negative as $T \rightarrow T_{\rm c}$ from above. At $T_{\rm c}$, $\partial D/ \partial U$ diverges. Therefore, by plotting the magnitude of $(\partial D/\partial U)^{-1}$ as a function of $T$, it will extrapolate to zero at $T_{\rm c}$, as shown by diamonds in Fig.~\ref{figS-Tc}(b). 

To find a lower bound for $T_{\rm c}$, we note that the hysteresis loops approaches zero as $T \rightarrow T_{\rm c}$ from below. Therefore, we extrapolate $T_{\rm c}$ by plotting the size of the coexistence region $U_{\rm c2}-U_{\rm c1}$ as a function of temperature, as shown by squares in Fig.~\ref{figS-Tc}(b).  

Operationally, we define $T_{\rm c}$ as the midpoint between the highest temperature where $D(U)$ shows hysteresis and the smallest temperature where $D(U)$ is continuous (see horizontal dashed lines in Fig.~\ref{figS-Tc}). Consequently, $U_{\rm c}$ is the midpoint between the value of $U_{\rm c2}$ corresponding to the highest temperature where we found hysteresis, and the value of $U$ where $D(U)$ has its largest slope. In summary, we obtain $U_{\rm c} \approx 5.90 \pm 0.05$ and $T_{\rm c} \approx 0.06 \pm 0.005$. 

Our estimate for the location of the Mott endpoint improves previous estimates with $2\times 2$ plaquette DMFT (Refs.~\onlinecite{phk, sht2}), and is close with those obtained with other methods: $4\times 4$ DCA gives $U_{\rm c} = 6.53$~\cite{vanLoon:2018} and dual fermion approach gives $U_{\rm c} = 6.64$~\cite{vanLoon:2018}.

\subsection{Single occupancy and charge compressibility}

The $U$-driven Mott transition at $n=1$ can also be revealed by the behavior of the occupation $n$ as a function of $\mu$ for different values of $U$ and $T$. As shown in Fig.~\ref{figS-n}(a),(b), the shape of $n(\mu)$ differs below and above $U_c$: for $U<U_{\rm c}$, $n(\mu)$ monotonically increases with increasing $\mu$, indicating metallic behavior from empty band ($n=0$) all the way until half-filled band, at $n=1$, or $\mu=U/2$. On the other hand, for $U>U_{\rm c}(T)$, $n(\mu)$ develops a plateau at $n=1$, signalling that the half-filled system is a Mott insulator. 

For $T<T_{\rm c}$ and within the coexistence region $U_{c1}(T)<U<U_{c2}(T)$ [see Fig.~\ref{figS-n}(c)] both a metallic and an insulating solution can be stabilized. As a result, close to $\mu=U/2$ two possible profiles of $n(\mu)$ coexist: a monotonically increasing function of $U$ (red up triangles) coexists with a flat curve at $n=1$ denoting Mott plateau (blue down triangles). In this article we confine our interest to the metal-insulator  transition driven by $U$ at half-filling, so the metal-insulator transition driven by doping (or, equivalently, by chemical potential) is not considered here (for a detailed discussion with the same methodology, see Refs.~\onlinecite{sht, sht2, ssht}).

\begin{figure}[hb!]
\centering{
\includegraphics[width=1\linewidth]{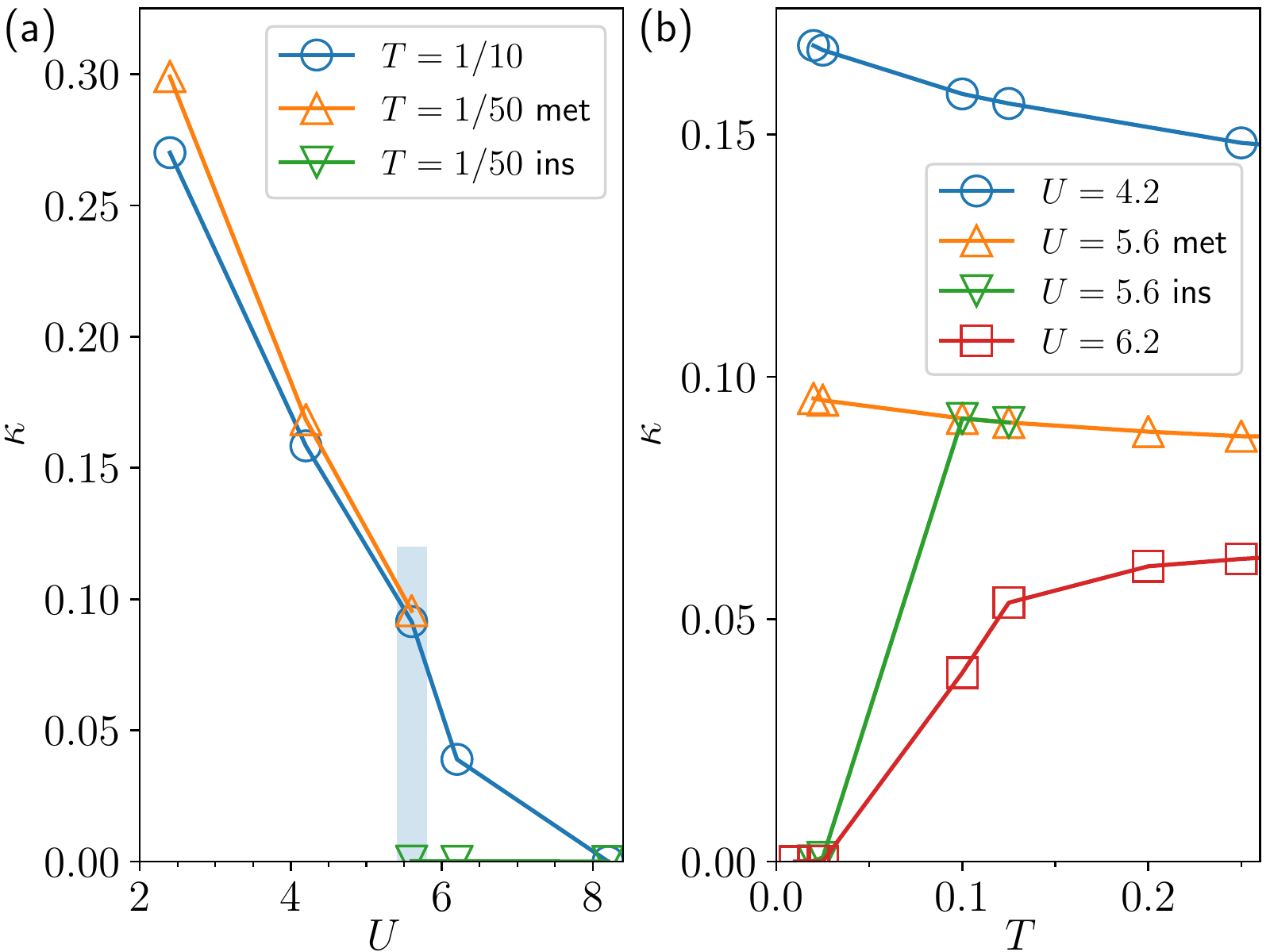}
}
\caption{(a) Isothermal charge compressibility $\kappa=(1/n^2) (dn/d\mu)_T$ at $n=1$ as a function of interaction strength $U$ and for $T$ above and below $T_c$, $T=1/10$ and $T=1/50$, respectively. Shaded area indicates the coexistence between a metal and an insulator, characterized respectively by finite and zero charge compressibility. (b) $\kappa$ versus temperature, for different values of $U$. For $U=5.6$ and low temperature, two solutions coexist. All lines are guides to the eye. 
}
\label{figS-kappa}
\end{figure}
\begin{figure*}
\centering{
\includegraphics[width=1\linewidth]{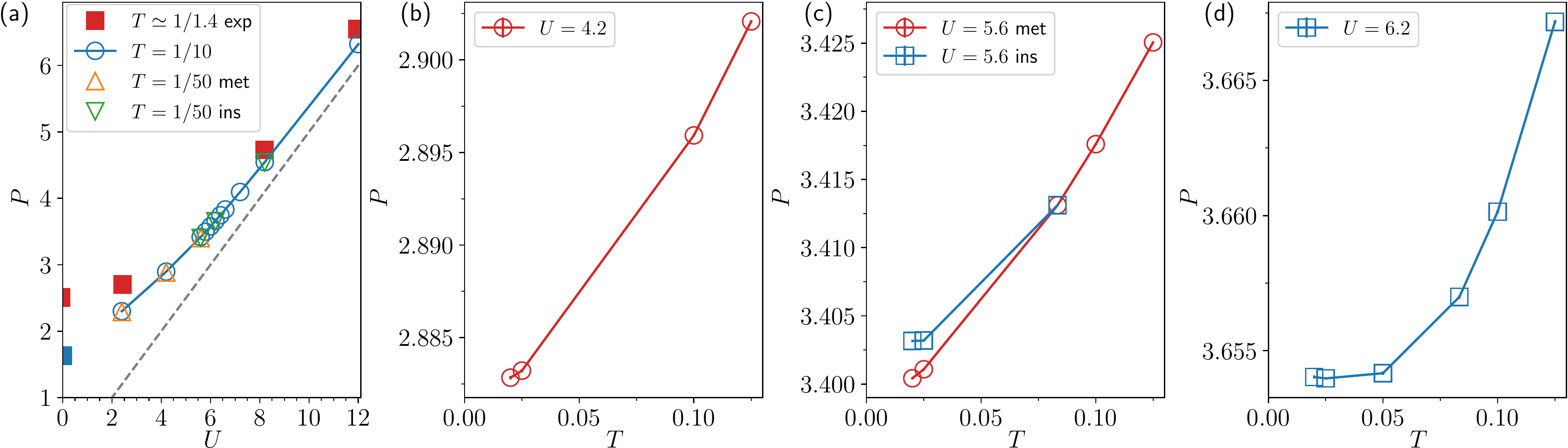}
}
\caption{(a) Pressure $P$ versus interaction strength $U$ at $n=1$ for different temperatures. Open symbols have been obtained using Eq.~(\ref{eq:p}). Integration has been performed using composite trapezoidal rule. Error bars, most of them invisible on this scale, indicate three standard deviations. Dashed line is the asymptotic behavior $P=U/2a$ in the limit $U\rightarrow \infty$, $T\rightarrow 0$. Filled red squares are experimental data for ultracold atoms in Ref.~\onlinecite{Cocchi:PRX2017} at $T\approx 1/1.4$. (b),(c),(d) Pressure $P$ versus temperature $T$ at $n=1$ for different values of $U$: $U=4.2<U_{\rm c}$ [panel (b)], $U=6.2>U_{\rm c}$ [panel (d)], and $U=5.6$ [panel (c)], which, at low temperatures, lies within the coexistence region. In that region, two solutions characterized by two distinct values of pressure coexist, with $P_{\rm met} < P_{\rm ins}$.  Note that for $U=6.2$, $P(T)$ show activated behavior and it is increasing with increasing $T$ within error bars.
}
\label{figS-P}
\end{figure*}

The slope of $n(\mu)$ is proportional to the charge compressibility $\kappa=(1/n^2) (dn/d\mu)_T$. Figure~\ref{figS-kappa}(a) shows $\kappa$ at $n=1$ as a function of $U$ for $T=1/10$ and $T=1/50$. As expected, the metallic state is compressible whereas the Mott insulator is incompressible. $\kappa$ decreases with increasing $U$ and for $T<T_{\rm c}$ shows a sudden jump at the Mott transition (see shaded region). Fig.~\ref{figS-kappa}(b) shows the temperature dependence of $\kappa$. At the low temperatures considered here, for $U<U_{\rm c}$, $\kappa$ monotonically decreases with increasing $T$, whereas for $U>U_{\rm c}$, $\kappa$ increases with increasing $T$, indicating thermal activation of electrons across the Mott gap. Compressibility has recently been measured in ultracold atom experiments~\cite{Cocchi:PRL2016} for the two-dimensional Hubbard model. 

Clearly, to calculate the isothermal charge compressibility at $n=1$, one does not need knowledge of $n(\mu)$ from empty to half-filled band: few points close to $\mu=U/2$ suffice. However, as we shall see in the next section, knowledge of $n(\mu)$ from empty to half-filled band allows us to obtain, using the Gibbs-Duhem relation, pressure, free energy and entropy across the Mott transition at $n=1$ and its precursor Widom line.

\section{Gibbs-Duhem relation}
\label{sec:GibbsDuhem}

In this section we exploit the Gibbs-Duhem relation to find pressure and entropy for the Mott transition and its supercritical crossover. These two thermodynamic quantities are found from knowledge of $n(\mu)$ from empty to half-filled band. This approach has been motivated by recent experiments with  ultracold atoms on the 2D Hubbard model~\cite{Cocchi:PRX2017}. It is extremely resourceful, yet computationally very expensive and thus not much applied. To our knowledge this method to find pressure or entropy has been used only for the attractive 2D Hubbard model~\cite{Anderson:PRL2015} and, with the two-particle-self-consistent approach~\cite{Vilk:1997}, for the metallic part of the 2D Hubbard model~\cite{Roy:2002}.

\subsection{Pressure}
\label{subsec:Pressure}

The Gibbs-Duhem relation is
\begin{align}
sdT -adP +nd\mu=0, 
\label{eqS:GD}
\end{align}
where $s$ is the entropy per particle, $a$ the surface per particle and $P$ the pressure. At constant $T$ and $U$ it becomes $nd\mu=adP$, from which one can extract $P$ by integrating $n(\mu)$ from $0$ (empty band) to $1$ (half-filled band), 
\begin{equation}
P(T)_U = \frac{1}{\emph{a}} \int_{-\infty}^{U/2} n(\mu, T) d\mu.
\label{eq:p}
\end{equation}

This step is computationally demanding, and to the best of our knowledge it has not been attempted so far within cluster extensions of DMFT, nor with single-site DMFT: in order to capture subtle variations of the occupation $n(\mu)_T$, careful scans in steps of $\mu$ ranging from $0.2$ down to $0.0025$ were performed~(see for example Fig.~\ref{figS-n}). This means that a single value of the pressure requires a detailed knowledge of $n(\mu)$, which we typically attain with of order of 50-170 $\mu$ values for CDMFT calculated $n$. 
Despite the high computational cost, we computed 31 values of the pressure across the Mott transition and the associated Widom line, for a total of more than 2000 points in the space of parameters given by temperature, chemical potential and interaction strength. 

For the numerical integration of $n(\mu)$, we use the composite trapezoidal rule and a lower limit of integration $\mu_{\rm min}$ corresponding to $n(\mu_{\rm min}) \approx 0.002$. We have verified that other integration methods (Simpson's rule and Romberg method) give the same result up to the 5th digits. This suggests that the error coming from integration on a finite grid is negligible. For the error bars associated to each value of the pressure, we consider the statistical error (associated to the error on the $n(\mu$)) only. To compensate our neglect of the systematic error associated to the discretized integral, our error bars contain three standard deviations. Further accuracy checks are given in the next two subsections. 

Figure~\ref{figS-P} shows the pressure $P$ as a function of interaction strength $U$ at $n=1$, for $T=1/10> T_{\rm c}$ (blue open circles) and $T=1/50<T_{\rm c}$ (open triangles). $P$ increases with increasing $U$ and approaches the asymptotic behavior $P\rightarrow U/(2\emph{a})$ obtained in the limit $U\rightarrow \infty$, $T\rightarrow 0$ (dashed line). We find consistency with experimental data on ultracold atoms in Ref.~\onlinecite{Cocchi:PRX2017} (filled red squares). The slight deviation downward over the entire range of $U$ is caused by the higher temperature used in experiments. 

Indeed, $P(T)$ increases with increasing temperature. Figures~\ref{figS-P}(b),(c),(d) show $P$ as a function of temperature for three values of the interaction strength: $U=4.2<U_{\rm c}$, $U=5.6$ which, for $T<T_{\rm c}$, lies within the coexistence region, and $U=6.2>U_{\rm c}$. Despite the small variation of the pressure with T and the error bars, it is clear that $P(T)$ increases more rapidly with increasing T in the metal than in the Mott insulator. Indeed, in a Fermi liquid, one expects $P \propto a + bT^2$ with $a$ and $b$ constants, whereas in a Mott insulator one expects activated behavior $P \propto a+b\exp(-\Delta_g/T)$, with $\Delta_g$ the Mott gap. Figure~\ref{figS-P}(c) shows that within the coexistence region, two distinct values of pressure appear, with $P_{\rm met} < P_{\rm ins}$.

\subsection{Entropy}
\label{subsec:Entropy}

\begin{figure}
\centering{
\includegraphics[width=1\linewidth]{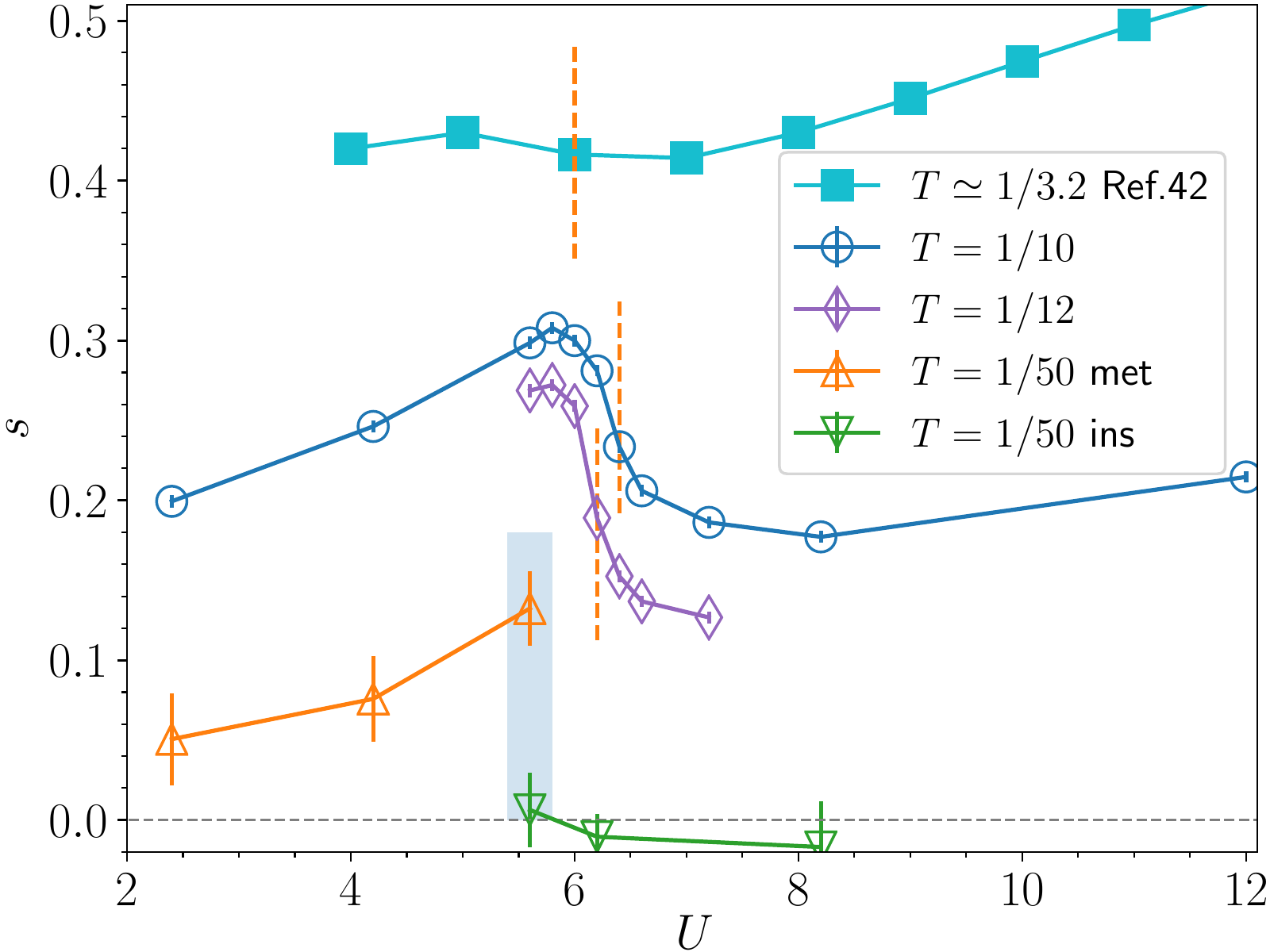}
}
\caption{Entropy per site $s$ versus interaction strength $U$ at $n=1$ for different temperatures $T$: $T=1/10, 1/12>T_{\rm c}$ and $T=1/50<T_{\rm c}$. Data have been obtained using Eq.~(\ref{eq:s}). Numerical derivatives have been performed using finite differences between two temperatures. We estimate $s$ at $T=1/10$ by taking finite differences between $T=1/8$ and $T/10$. We estimate $s$ at $T=1/12$ by taking finite differences between $T=1/10$ and $T=1/12$. We estimate $s$ at $T=1/50$ by taking finite differences between $T=1/40$ and $T=1/50$. Error bars indicate three standard deviations. Orange dashed vertical lines mark the inflection point in $s(U)_T$ above $T_{\rm c}$. Shaded area marks the coexistence between metal and insulator below $T_{\rm c}$. Since in the Mott insulator at low temperature the pressure $P(T)$ shows activated behavior [see Fig.~\ref{figS-P}(d)], the entropy is zero within error bars. Filled squares show high-temperature data of Ref.~\onlinecite{KhatamiPRA2011}. 
}
\label{figS-s1}
\end{figure}
The entropy per site $s=- {\rm Tr} [\rho \ln \rho]/N$, where $\rho$ is the density matrix and $N$ the number of sites, can be obtained from the Gibbs-Duhem relation as:
\begin{equation}\label{eq:s}
s = \emph{a} \left( {dP}/{dT} \right)_\mu.
\end{equation}
For the numerical derivative, we perform finite differences between two temperatures. 

Figure~\ref{figS-s1} shows the entropy $s$ as a function of interaction strength $U$ for three temperatures, $T=1/10, 1/12>T_{\rm c}$ and $T=1/50<T_{\rm c}$. 
Let us first focus on $T=1/10$, which is larger than, but not far from, $T_{\rm c}$: $s(U)_T$ exhibits non-monotonic behavior: at first, the entropy increases with increasing $U$ until it reaches a local maximum. The increase of entropy with $U$ coming from the metallic side is easy to understand since entropy is proportional to effective mass in Fermi-liquid theory and effective mass increases with interactions. 
Increasing $U$ further, $s(U)_T$ shows a sharp drop marked by an inflection point (vertical orange dashed line), followed by a shallow local minimum. 
With $U$ even larger, the entropy increases with $U$, asymptotically reaching $\ln 2$, as expected for localized independent spins, at $U\rightarrow \infty$ (not shown). 
The sharp decrease of $s$ with $U$ that precedes the increase towards $\ln 2$ should occur even in infinite-system calculations because the entropy from spin waves is inversely proportional to the square of the spin-wave velocity. That velocity increases with $U$ as long as the Mott transition occurs before the asymptotic Heisenberg regime, which is observed to be the case~\cite{phk, balzer, sht2}.  

The non-monotonic behavior of our data is compatible with numerical linked-cluster expansions results at higher temperature~\cite{KhatamiPRA2011} shown by filled squares in Fig.~\ref{figS-s1} for $T\approx 1/3.2$.  

\begin{figure*}
\centering{
\includegraphics[width=1\linewidth]{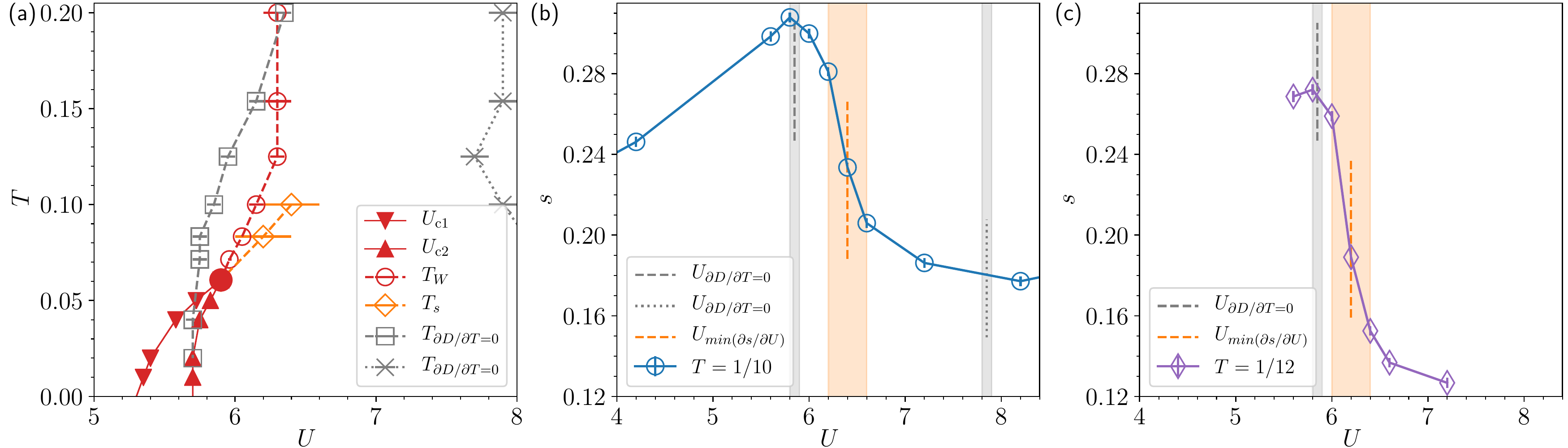}
}
\caption{(a) Temperature $T$ - interaction strength $U$ phase diagram of the 2D Hubbard model at $n=1$. Orange open diamonds denote the crossover line $T_s$ obtained as the loci of inflection points in the most rapid downward fall with $U$, as measured by the position of the local minima min$(\partial S/ \partial U)_T$. It emanates out of the Mott endpoint (full red circle) into the supercritical region. Red symbols are as in Fig.~\ref{figS-PhaseDiagram}: red triangles indicate the coexistence line and red open circles mark the Widom line $T_W$ as obtained from the loci of the inflections in $D(U)_T$. The gray squares (crosses) point to the position of the local maxima (local minima) of the entropy $s(U)_T$. They are determined using the Maxwell relation $(\partial s / \partial U)_{T,n} = -(\partial D / \partial T)_{n,U}$. At each temperature, the crossover line extracted from the loci of inflection point (orange diamonds) occurs between the lines with gray symbols. 
(b),(c) Entropy per site $s$ versus $U$ for $T=1/10$ [panel (b)] and $T=1/12$ [panel (c)]. Vertical orange dashed line marks the position of the inflection, near the sharpest drop with $U$. The position of the inflection is shown in panel (a) with orange diamonds. Vertical gray dashed (dotted) line denotes the position of the local maximum (local minimum) in $s(U)_T$ as calculated by the Maxwell relation. The good agreement with the actual data (open symbols) provides a consistency check for our calculation. The position of the local maximum (local minimum) at each temperature is shown in panel (a) with gray squares (gray crosses).
}
\label{figS-s2}
\end{figure*}

The key feature of $s(U)_T$ is the inflection point marking the most rapid decrease of the entropy with $U$ (vertical red dashed lines): for example, at $T=1/10$, with $U$ increasing from 6 to 7, the entropy  drops by almost one-third. Furthermore, the slope associated to this sharp drop becomes steeper with progressively decreasing $T$, as demonstrated by our data at $T=1/12$ (violet diamonds). Figure~\ref{figS-cfr}(e) indeed shows that the minimum of $(\partial s/ \partial U)_T$ increases in magnitude in going from $T=1/10$ to $T=1/12$. Thus we expect that at $T_{\rm c}$, the inflection turns into a infinite slope. This is indeed the case: at $T_{\rm c}$, it is straightforward to demonstrate that $s(U)$ scales as $-{\rm sgn}(U-U_c)|U-U_c|^{1/\delta}$ and thus has infinite slope since $\delta>1$ (see Appendix~\ref{sec:CriticalS}). 

By tracking the position of the inflection points of the entropy $s(U)_T$, we can thus define a crossover line in the $T-U$ phase diagram, in complete analogy to our analysis of the inflection points of the double occupancy described in Sec.~\ref{sec:Docc}. This crossover is shown in Figure~\ref{figS-s2}(a) (orange diamonds). Figures~\ref{figS-s2}(b),(c) zoom-in on the behavior of $s(U)_T$ close to $U_{\rm c}$ for $T=1/10$ and $T=1/12$ respectively. Similarly to the Widom line, the crossover marking the sharpest variation of entropy with $U$ evolves into the Mott critical endpoint. It closely follows the Widom line and indeed we expect that asymptotically close to $T_{\rm c}$ all these crossovers merge on the same line. Due to the high computation cost, we were able to obtain the crossover in the entropy only at two temperatures, $T=1/10$ and $T=1/12$. However, an inflection point still occurs in the high-temperature data of Ref.~\onlinecite{KhatamiPRA2011} (see orange vertical dashed line in Fig.~\ref{figS-s1}), suggesting that this crossover may persist up to quite high temperature. The sharp crossover in the entropy emerging from the Mott endpoint into the supercritical region is one of the key findings of our work.

Below $T_{\rm c}$ the crossover in the entropy evolves into a first-order transition. For $T=1/50<T_{\rm c}$, shown by triangles in Fig.~\ref{figS-s1}, $s(U)$ is dramatically reduced in the Mott insulator because charge excitations are gapped while spin fluctuations are reduced due to short-range singlet formation~\cite{phk, sht, sht2}. The collapse to zero of the entropy differs from single-site DMFT, where the Mott insulator has $\ln 2$ ground state entropy~\cite{rmp, phk}. Within the coexistence region, the entropy is discontinuous, with $s_{\rm ins} < s_{\rm met}$, resulting in the latent heat $\ell = T(s_{\rm met}-s_{\rm ins})$. Heat must be added to melt the insulator into the metal. We find $\ell \approx 0.0025$ for $U=5.6$ and $T=1/50$. 

The Clausius-Clapeyron equation, $dT/dU=(D_{\rm ins}-D_{\rm met})/(s_{\rm ins}-s_{\rm met}) = (D_{\rm ins}-D_{\rm met})/(\ell T)$ relates latent heat to the difference in double occupancy and to the slope of the coexistence curve $dT/dU$. The metallic phase has larger entropy and larger double occupancy than the insulating phase, implying a positive slope for the coexistence curve, in agreement with our $T-U$ phase diagram (see e.g. Fig.~\ref{figS-s2}(a) and Fig.~\ref{figS-PhaseDiagram}). Previous works~\cite{phk} inferred from the positive slope of the coexistence curve on the $T-U$ phase diagram that the Mott insulator has lower entropy than the metal. By calculating the entropy, our contribution is to quantify the discontinuity of the entropy across the Mott transition. Furthermore, we note that as $T\rightarrow 0$, the slope of the first-order transition, $dT/dU$, becomes vertical. This follows from the Clausius-Clapeyron relation and the fact that the transition at $T=0$ is between two states with same $s=0$ entropy. Our result on the infinite slope of $dT/dU$ at $T=0$ corrects what has been previously suggested~\cite{phk, balzer}.

We end this section with a remark that also serves as a further consistency check on the behavior of the entropy with $U$. For $T>T_{\rm c}$, $s(U)_T$ shows two extrema: a local maximum for $U<U_{\rm c}$ and a local minimum for $U>U_{\rm c}$ [see e.g. Fig.~\ref{figS-s2}(b),(c)]. A Maxwell relation prescribes that the extrema of the entropy can also be determined by the crossing of the isotherms $\partial D / \partial T = 0$. The proof works as follows: $d(e-Ts)=-sdT-Pda+\mu dn+DdU$ implies that $\partial D/\partial T|_{a,n,U}=-\partial s/\partial U|_{a,n,T}=0$~\cite{rmp,WernerAdiabCooling2005,Dare:2007,Paiva_Scalettar_Randeria_Trivedi_2010}. The gray vertical lines in Fig.~\ref{figS-s2}(b),(c) indicate the value of $U$ at which the two extrema computed using the Maxwell relation occur. The agreement with the position of the local maximum of $s(U)_T$ calculated using the Gibbs-Duhem relation is excellent for both $T=1/10$ and $T=1/12$. To save computing time, the consistency between the two methods in the case of the minimum has been verified only for $T=1/10$. The loci of the entropy extrema are plotted in the $T-U$ phase diagram of Fig.~\ref{figS-s2}(a) (gray squares and crosses). Note that the crossover emerging from the Mott endpoint (orange diamonds) is in between the loci of entropy extrema.

\subsection{Accuracy check: kinetic energy}
\label{subsec:Ekin}

\begin{figure}
\centering{
\includegraphics[width=1\linewidth]{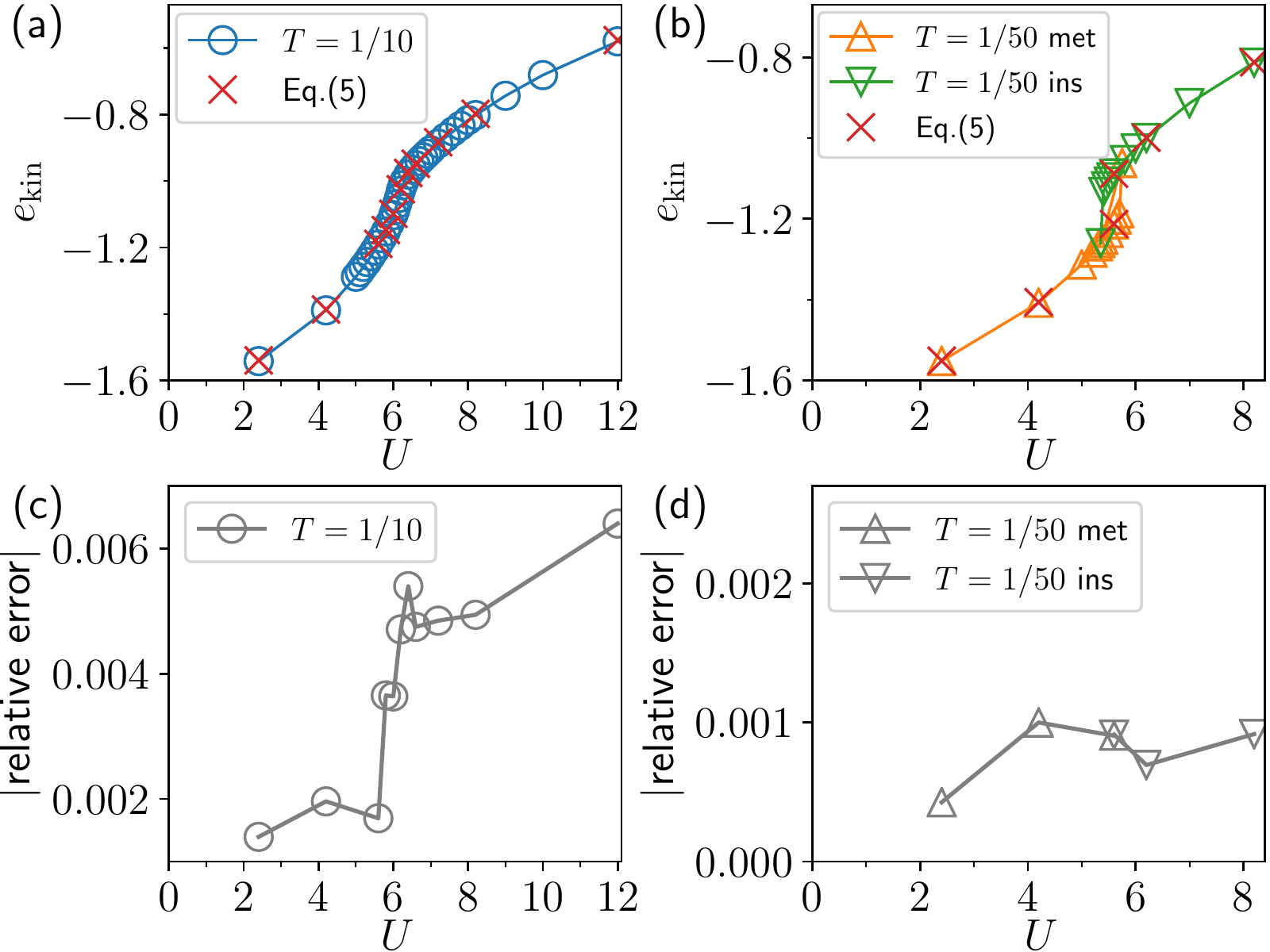}
}
\caption{Kinetic energy per site $e_{\rm kin}$ versus $U$ for $T=1/10> T_{\rm c}$ [panel (a)] and for $T=1/50 < T_{\rm c}$ [panel (b)]. We calculate $e_{\rm kin}$ with two different methods: within the CT-HYB impurity solver (open symbols) and using Eq.~(\ref{eq:Ekin}) (X symbols), $e_{\rm kin} = Ts +\mu n -Pa-UD$. 
Relative error as a function of $U$ for $T=1/10$ [panel (c)] and for $T=1/50$ [panel (d)]. The overall agreement provides a strong consistency check for our determination of the pressure  $P$ and the entropy $s$. 
}
\label{figS-KE}
\end{figure}

The subtle variations of pressure and of entropy with $U$ and $T$ are one of our central results. We have already discussed the intrinsic checks that we have performed. In this subsection we discuss one of the most stringent consistency checks in our calculations: the calculation of the kinetic energy. From the entropy and the pressure, one can calculate the kinetic energy per site from the Gibbs-Duhem result
\begin{equation}
e_{\rm kin} = Ts +\mu n -Pa-UD. 
\label{eq:Ekin}
\end{equation}
Figures~\ref{figS-KE}(a),(b) show $e_{\rm kin}$ as a function of $U$ for two temperatures, $T=1/10>T_{\rm c}$ and $T=1/100<T_{\rm c}$ calculated using Eq.~(\ref{eq:Ekin}) (X symbols). Alternatively, the kinetic energy can be extracted with high accuracy directly within the CT-HYB impurity solver (open symbols): Ref.~\onlinecite{LorenzoSC}  demonstrated that $e_{\rm kin}$ is the sum of two terms, a contribution related to the average expansion order term, plus a term coming from the cluster part. Figures~\ref{figS-KE}(c),(d) show the relative error between the two methods as a function of $U$ both above and below $T_{\rm c}$: the overall relative error is smaller than $1\%$, therefore implying excellent internal consistency.

\section{Thermodynamic stability}
\label{sec:Stability}

We have all that is needed to compute thermodynamic potentials. This is discussed in the first subsection below. Concavity of the grand potential is linked to thermodynamic {\it stability}. Stability criteria give a more fundamental and unifying understanding of thermodynamics. In the following two subsections, we thus study the local stability, i.e. stability under small perturbations, and then the global stability, i.e. which phase minimizes the grand potential.

\subsection{Grand potential}
\label{sec:FreeEnergy}

\begin{figure*}
\centering{
\includegraphics[width=1\linewidth]{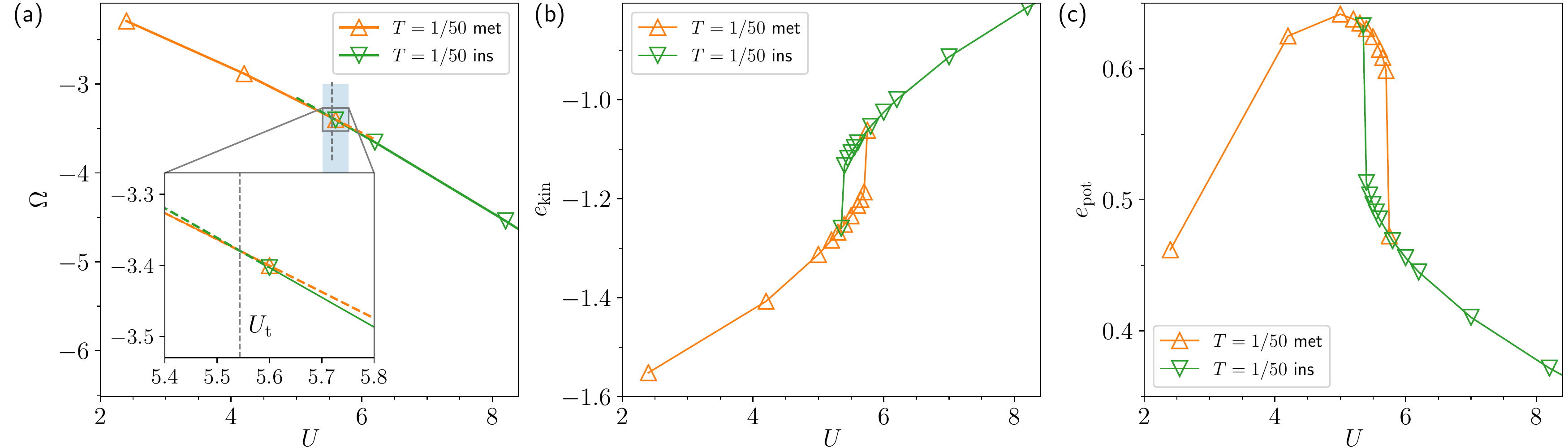}
}
\caption{(a) Grand potential $\Omega=-P=E_{\rm kin}+DU-Ts-\mu n$ for $T=1/50 < T_{\rm c}$. Lines with up and down triangles indicate the metallic and insulating solutions, respectively. Shaded area indicates the coexistence region between metal and insulator. The inset zooms on this region, to show that $\Omega_{\rm ins} < \Omega_{\rm met}$ for $U=5.6$. By linear extrapolation of the nearest points just below and just above $U=5.6$ (dashed orange and green lines, respectively), we can estimate the value of $U$ where the grand potentials cross, $U_t$ (vertical gray dashed line). We have repeated a similar analysis for $T=1/40$ (not shown). The location of $U_t$ for these temperatures is shown with red crosses on the $T-U$ phase diagram in Fig.~\ref{figS-PhaseDiagram}.  (b),(c) Kinetic energy $e_{\rm kin}$ and potential energy $e_{\rm pot}=UD$ versus $U$ for the same temperature as panel (a). Within the coexistence region, the kinetic and potential energy differences are much larger than the grand-potential difference.   
}
\label{figS-F}
\end{figure*}

From Eq.~(\ref{eq:Ekin}), $e_{\rm kin} = Ts +\mu n -Pa-UD$, it follows that the grand potential per site $\Omega=e_{\rm kin}+UD -Ts-\mu n$ (usually extracted from the partition function in the grand-canonial ensemble), is equal in magnitude and opposite in sign to the pressure, $\Omega=-Pa$. Since $P$ is a convex function of $T$ and $U$ (see Fig.~\ref{figS-P}), as expected $\Omega$ is a concave function of $T$ and $U$. Figure~\ref{figS-F} shows $\Omega(U)$ for $T=1/50 < T_{\rm c}$ [panel (a)] along with the kinetic energy $e_{\rm kin}(U)$ [panel (b)] and potential energy $e_{\rm pot}(U)=UD(U)$ [panel (c)].

\subsection{Local stability}
\label{sec:LocalStability}

The Mott insulating phase and the metallic phase separated by the Mott transition are locally stable. The proof follows our Ref.~\onlinecite{sht2} and is shown here for completeness. 

In the grand-canonical ensemble mentioned above, the natural thermodynamic variables are temperature, chemical potential and volume (here area). We also include interaction strength $U$ since we are at fixed filling and volume, wishing to study stability in the $T-U$ plane. The corresponding conjugate variables can be deduced from 
\begin{align}
d\Omega(T,\mu,a,U)&=d(e_{kin}+UD-Ts-\mu n)\\
                  &=-sdT - nd\mu -Pda + DdU.
\label{eq:dE}
\end{align}

If instead of controlling $\mu$ we control $n$, then the appropriate Legendre transform leads us to the Helmholtz free energy: 
\begin{align}
df &= d(e_{kin}+UD-Ts) \\
   &= -sdT + \mu dn -Pda + DdU.
\label{eq:defdO}
\end{align}
Taking area and filling fixed from now on, we can focus on $df = -sdT + DdU.$ We thus have, dropping constant $n$ and constant $a$ symbols, 
\begin{align}
\left(\frac{\partial f}{\partial T}\right)_{U} = -s \; ; \qquad \left(\frac{\partial f}{\partial U}\right)_{T} = D.
\end{align}

Local stability requires  $d^2f<0$. In matrix notation, this reads:
\begin{equation}
\begin{split}
d^2 f  = \begin{pmatrix} dT & dU \end{pmatrix}
             \begin{pmatrix} \left(\frac{\partial^2 f}{\partial T^2}\right)_{U} & \left(\frac{\partial^2 f}{\partial T \partial U}\right) \\  
             \left(\frac{\partial^2 f}{\partial T \partial U}\right) & \left(\frac{\partial^2 f}{\partial U^2}\right)_{T} \end{pmatrix}
             \begin{pmatrix} dT \\ dU \end{pmatrix} < 0.
\end{split}
\end{equation}
At constant $T$, this inequality becomes: 
\begin{equation}
\left(\frac{\partial^2 f}{\partial U^2}\right)_{T} = \left(\frac{\partial D}{\partial U}\right)_{T} < 0,
\end{equation}
which is satisfied by our results, including metastable phases, as shown in Fig.~\ref{figS-docc}. Note that
\begin{equation}
\left(\frac{\partial^2 f}{\partial T^2}\right)_{U} = -\left(\frac{\partial s}{\partial T}\right)_{U} < 0 \; ; \; \left(\frac{\partial^2 f}{\partial T \partial U} \right) =-\left(\frac{\partial D}{\partial T} \right)_{U}. 
\end{equation}
The sign of the mixed derivative is arbitrary, as long as the determinant of the matrix for $d^2f<0$ is positive, namely 
\begin{equation}
\left(\frac{\partial^2 f}{\partial T^2}\right)_{U}  \left(\frac{\partial^2 f}{\partial U^2}\right)_{T} -\left(\frac{\partial^2 f}{\partial T \partial U} \right)^2 > 0.
\label{eq:stability}
\end{equation}
At the critical endpoint the free energy is no longer analytic. The second derivatives, taken from either directions approaching the critical point, become negative infinity, namely $\left({\partial^2 f}/{\partial U^2}\right)_{T}|_{(T_{\rm c},U_{\rm c}^\pm)} \rightarrow -\infty$ and $\left({\partial^2 f}/{\partial T^2}\right)_{U}|_{(T_{\rm c}^+,U_{\rm c})} \rightarrow -\infty$.

Above $T_{\rm c}$, at each point of the phase diagram the free energy is locally stable and uniquely determined by $T$ and $U$. Below $T_{\rm c}$, in the coexistence region, there are two locally stable phases corresponding to the metallic and insulating phases, as shown in Fig.~\ref{figS-F}(a). The preferred phase is the one that minimizes the grand potential. This is the condition of global stability discussed below.

\subsection{Global stability}
\label{sec:GlobalStability}

A global stability analysis tells us which phase minimizes the grand potential (or equivalently the Helmholtz free energy since adding $\mu n$ to the grand potential of the two phases cannot change their intersection in the $T-U$ plane at constant $n$). Within the coexistence region, the grand potential $\Omega$ has two values, corresponding to the existence of a metallic phase and an insulating phase. The first-order transition occurs where the grand potentials cross. 

Figure~\ref{figS-F}(a) and its inset, zooming on the coexistence region, show the grand potential at $T=1/50< T_{\rm c}$. At $U=5.6$, which is the only point we have within the coexistence region, we found that $\Omega_{\rm ins} < \Omega_{\rm met}$. To estimate where the crossing of the grand potentials occur, we linearly extrapolate the nearest points just below and just above $U=5.6$ (dashed orange and green line, respectively). We find that the grand potentials for the metallic and the insulating solutions cross at $U_{t} \approx 5.55$ (gray dashed vertical line). For $U>U_t$ the insulating solution is globally stable and the metallic solution is only metastable (at $U=5.6$ we indeed find $\Omega_{\rm ins} < \Omega_{\rm met}$), while for $U<U_t$ the opposite occurs. We conducted this analysis for two temperatures, $T=1/50$ [Fig.~\ref{figS-F}(a)] and $T=1/40$ (not shown). The loci of points formed by $U_t(T=1/50)$ and $U_t(T=1/40)$ allow us to obtain an estimate for the thermodynamic first-order transition line $U_t(T)$ in the $T-U$ phase diagram of Fig.~\ref{figS-PhaseDiagram} where red crosses within the coexistence region show the values of $U_t(T)$. 

Two remarks are in order. First, given the few points at which we can evaluate the $\Omega$, our $U_t(T)$ curve is only a crude estimate of the first-order transition line. However, this has not been attempted before. Given that $U(T)$ must have a vertical tangent for $T\rightarrow 0$ (see the discussion at the end of subsection \ref{subsec:Entropy}), our results are compatible with a first-order transition at $T=0$~\cite{balzer}. 

Second, at the first-order transition the discontinuity in grand potential, $\Delta \Omega = |\Omega_{\rm ins}-\Omega_{\rm met} |$, is much smaller than the discontinuity in the kinetic energy $\Delta e_{\rm kin} = |(e_{\rm kin})_{\rm ins} - (e_{\rm kin})_{\rm met}|$ and in potential energy $\Delta e_{\rm pot} = |(e_{\rm pot})_{\rm ins} - (e_{\rm pot})_{\rm met}|$. 
This implies that the potential energy loss due to localization is almost perfectly compensated by the kinetic energy gain due to delocalisation, as already noticed within the single-site DMFT solution of the half-filled Hubbard model~\cite{rmp, mkz}.  
At fixed chemical potential and area, $\Delta \Omega = |\Omega_{\rm ins}-\Omega_{\rm met} |$ controls the critical temperature $T_c$~\cite{rmp, Antoine:Review2004} because in that case, $\Delta \Omega = \Delta e - T \Delta s$, so that $T_c \approx \Delta E / \Delta S$, helping us understand why $T_c$ is much smaller than the bare $t$.

\begin{figure}
\centering{
\includegraphics[width=1\linewidth]{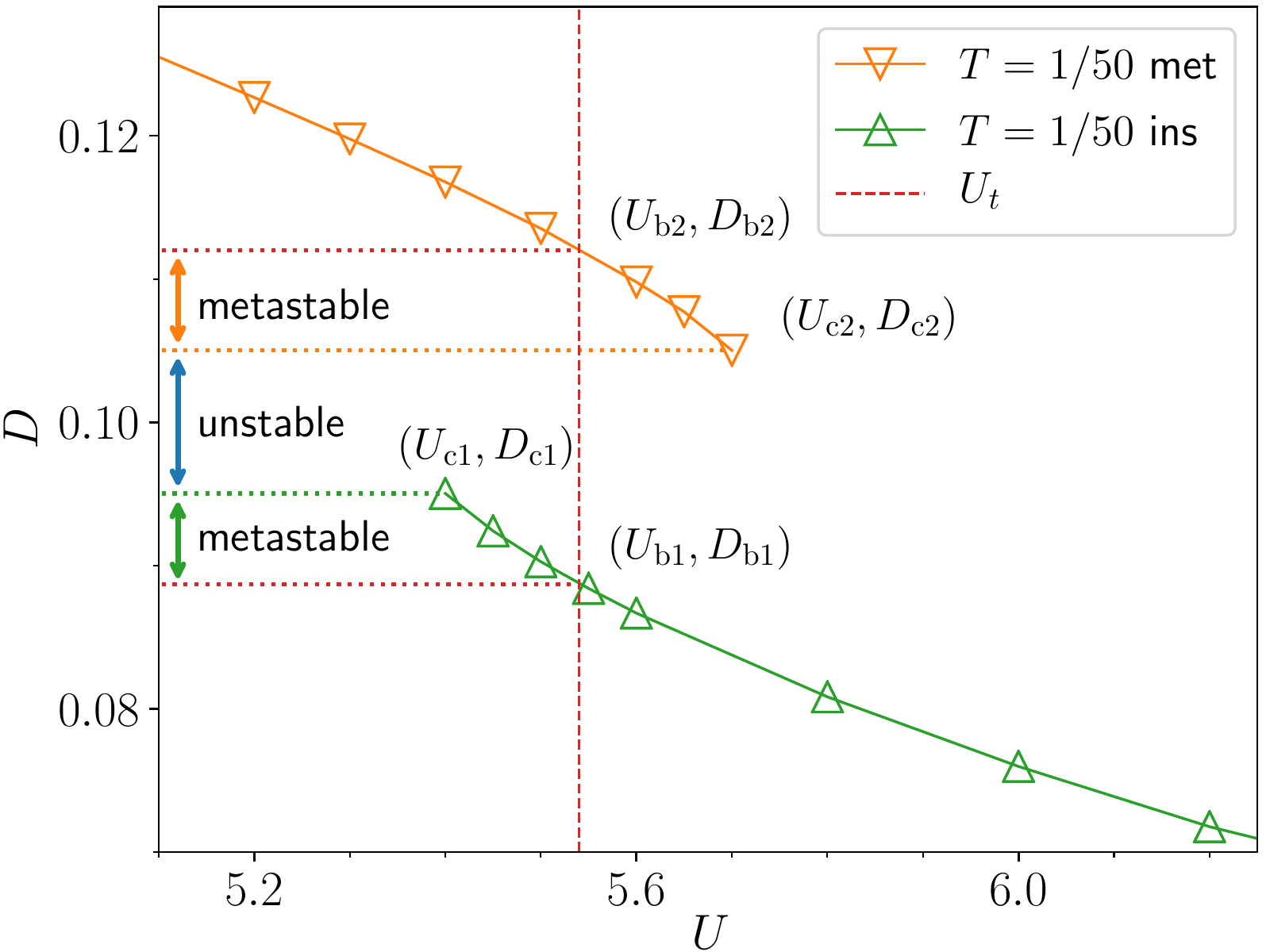}
}
\caption{Double occupancy $D$ versus $U$ at $T=1/50 < T_c$. Spinodal points and binodal points are indicated. Vertical arrows denote the metastable regions and the unstable region. 
}
\label{figS-m}
\end{figure}

Knowledge of the first-order transition line $U_t(T)$ allows us to consider issues of nucleation and spinodal decomposition. In the $T-U$ plane, we have just shown that the transition line $U_t(T)$ divides in two the coexistence region bounded by the spinodals $U_{\rm c1}(T)$ and $U_{\rm c2}(T)$: between $U_{\rm c1}(T)$ and $U_t(T)$ the metallic phase is globally stable and the insulating phase is metastable, whereas between $U_t(T)$ and $U_{\rm c2}(T)$ the insulating phase is globally stable and the metallic phase metastable. 
To proceed further, let us consider the double occupancy $D(U)$ shown in Fig.~\ref{figS-m} at $T=1/50$. The spinodal points $(U_{\rm c1}, D_{\rm c1})$, $(U_{\rm c2}, D_{\rm c2})$ indicate where the insulating and metallic solutions cease to exist, respectively. The region between $D_{\rm c2}$ and $D_{\rm c1}$ is {\it unstable}. 
The intercepts between $D(U)_T$ and the value of the thermodynamic transition $U_t$ (vertical dashed line) gives the binodal points $(U_{\rm b1}, D_{\rm b1})$, $(U_{\rm b2}, D_{\rm b2})$. The region between $D_{\rm b2}$ and $D_{\rm c2}$ and between $D_{\rm c1}$ and $D_{\rm b1}$ are {\it metastable}. 
By repeating the same analysis at different temperatures one could obtain a $D-U$ phase diagram with two spinodal lines surrounding the unstable region. The regions between binodals and spinodals are metastable. Binodals and spinodals emerge out of the critical endpoint $(U_{\rm c}, D_{\rm c})$. 

To progress further, let us recall the analogy between the Mott transition in electronic fluids and the liquid-gas transition in classical fluids~\cite{CastellaniAnalogy, Jaraman:1970, gabiEPJB, lange, mck, limelette,  Antoine:Review2004}: the metal corresponds to a high density liquid with a large number of double occupancies and holes so that electrons can delocalize throughout the lattice, whereas the Mott insulator corresponds to a low density incompressible gas with few double occupancies and holes so that electrons are localized. 
Now, in the metastable region, the electronic fluid is not thermodynamically stable and the other stable phase is triggered by nucleation or cavitation: for instance, thermal fluctuations may create droplets (or bubbles) of the other phase, which lower the grand potential energy, and which grow to nucleate the other phase. 
In the unstable region, the system phase separates through the so-called spinodal decomposition mechanism. 

The nucleation mechanism and spinodal decomposition mechanism can provide a framework for recent experimental studies focusing on the processes by which a Mott insulator transforms into a metal. For instance, textured states are observed across the Mott transition in V$_2$O$_3$~\cite{McLeod:2017} and VO$_2$~\cite{Qazilbash:2007}. Ultrafast dynamics can trigger nucleations of metallic droplets at the transition~\cite{Ronchi:arXiv2018, Singer:2018, Abreu:2015, OCallahan:2016}.

\section{Information-theoretic description}
\label{sec:Entanglement}

In the last two sections we have obtained a thermodynamic and statistical description of the Mott transition. In recent years, information-theoretic methods have provided new tools to analyse phase transitions in correlated many-body quantum systems~\cite{amicoRMP2008, EisertRMP}. It is therefore interesting to study the Mott transition with these tools.

In information theory, a cardinal concept is the one of {\it correlation} among parts of a system, \emph{i.e.} the information contained in one part of the system about the other parts. Two parts of a system are correlated if our ignorance (entropy) about one part can be decreased by observing the other part~\cite{CoverInformation, watrous2018}. 
Correlations in many-body quantum systems clearly play a role in observable properties, since collective phenomena that emerge at a macroscopic scale are not a simple sum of microscopic properties~\cite{AndersonMore}. 
By bringing together the theory of phase transitions and critical phenomena to quantum information theory, one goal is to gain a novel perspective on the correlations underlying phase changes in many-body quantum systems. 

This is a rich research programme, with ramifications in many areas of physics. For the purpose of our discussion, we confine ourselves on two issues about correlations at a phase transition. 

A first issue deals with what type of correlations one is looking for at a phase transition between many-body quantum systems. 
From the point of view of information theory, correlations in classical mechanics can only arise due to lack of knowledge about the system. Indeed, when a complete microscopic description of the system is available, then there is simply nothing to be learned about a part of the system by observing another part, hence no correlations. 
However, when we do not have full knowledge of the classical system, a probabilistic description becomes necessary and correlation functions become non-trivial. 
This contrasts with quantum mechanics where a complete description of the whole does not imply complete knowledge of the parts. This is the hallmark of entanglement, one of the most distinguished signature of quantum effects. Complete knowledge of the whole system is possible at zero temperature where the global state is pure. In this limit, any correlations in the system can thus be attributed to entanglement, and a faithful measure of entanglement is provided by the local entanglement entropy. But at finite temperature, only a statistical description of the whole system is available in the form of a density matrix. In that case, correlations can arise from both quantum fluctuations (quantum correlations) or thermal fluctuations (classical correlations). 
In addition, local thermal fluctuations contribute to the local entropy but do not contribute to correlations, which motivates the use of a more refined measure of correlations, such as the mutual information~\cite{CoverInformation, watrous2018}.

A second issue deals with the role of correlations at a phase transition in many-body quantum systems. Role of the correlations here indicates two main aspects: what is the behavior of the correlation measures as a function of the tuning parameters of the phase transition; and what is the structure of the distribution of correlations at a phase transition. Seminal works~\cite{OsterlohNature2002, OsbornePRA2002} on the relation between entanglement and quantum phase transition in spin systems showed that entanglement measures can indeed detect a quantum phase transition. These works opened up a way to many studies characterising quantum phase transitions with entanglement in correlated systems of spins, bosons and fermions: suitable entanglement measures can pick up the location of quantum phase transitions, can identify their first or second order character, and the associated critical exponents. Furthermore, one route to access the structure of the distribution of correlations at phase transitions is to do scaling analysis, i.e. to analyse how correlation measures scale as a function of distance and the number of sites~\cite{EisertRMP}. 

In the companion letter~\cite{walshSl}, we do a first step to characterize the Mott metal-insulator transition in the two-dimensional Hubbard model with information-theoretic tools. With respect to the first issue identified in the above discussion (i.e. what type of correlations we are looking for at the transition), we focus on two key measures of correlations, local entanglement entropy and mutual information. With respect to the second issue discussed above (i.e. what is the role of correlations at the transition), we confine ourselves on the behavior of entanglement entropy and mutual information as a function of the tuning parameters of the Mott transition, here interaction strength $U$ and temperature $T$. We showed that they characterize the first-order Mott transition: they detect the first-order nature of the transition by showing hysteretic behavior, they identify universality class of the Mott endpoint by showing critical scaling, and they pick up the crossover emanating from the endpoint in the supercritical region by showing sharp variations in marked by inflections. 
In the following two subsections, we present additional discussion of these two measures of correlations at the Mott transition, local entropy and total mutual information, respectively.

Main motivations for our study are twofold. First, up to now, most work focused on zero temperature, where only quantum correlations occur. The relation between entanglement measures and quantum phase transitions, both second-order~\cite{OsterlohNature2002, OsbornePRA2002, Vidal:2003, Vidal:2004, Wu:PRL2004, CamposPRA2006, Frerot_Roscilde_2016} and first-order~\cite{Bose:2002, Alcaraz:2004, Vidal:PRA2004, Wu:PRL2004}, has been explored in different many-body systems (for a review, see Ref.~\onlinecite{amicoRMP2008}). The relation between quantum phase transitions and correlation measures other than entanglement, has also been studied~\cite{Dillenschneider:2008, Sarandy:2009, Maziero:2010, Allegra:2011} (for reviews, see Ref.~\onlinecite{amicoRMP2008, EisertRMP, VedralRMPDiscord}). Fewer results have been obtained in the most difficult case of finite temperatures. In the finite temperature regime, it is the quantum mutual information that plays a role in quantifying the classical and quantum correlations~\cite{groisman2005, Melko:2010, Singh:2011, Kallin:2011, Wilms:2012}. Interesting open research directions are the study of finite-temperature continuous transitions~\cite{Wilms_Troyer_Verstraete_2011, Iaconis:2013} and the study of the finite temperature crossover emanating out of a quantum critical point~\cite{Amico_Patan_2007, Werlang:2010, Werlang:PRB2010, Gabbrielli:arXiv2018, Frerot:arXiv2018}. The Mott transition investigated in this work and in our companion letter presents the exciting possibility to study, within the same model, a first-order transition from its low temperature quantum limit, to its finite-temperature critical endpoint, and its associated crossover in the supercritical region. 
A second pressing motivation for our analysis of the Mott transition is that the entanglement properties are in general very elusive to measure experimentally, but very recent experiments with ultracold atoms~\cite{Cocchi:PRX2017, greinerNat2015} have removed this barrier, by measuring entanglement entropy and mutual information in 1D Bose-Hubbard model~\cite{greinerNat2015}, and 2D fermionic Hubbard model~\cite{Cocchi:PRX2017}. These groundbreaking works open up the avenue to experimentally detect quantum-information measures and call for theoretical work.

\subsection{Local entropy}

A key measure of entanglement for many-body quantum systems at $T=0$ is the entanglement entropy. 
The entanglement entropy is defined as $s_A=- {\rm Tr_A} [\rho_A \ln \rho_A]$, where the reduced density matrix $\rho_A$ is obtained by tracing the density matrix of the whole system over the remaining part $\overline{A}$, ($\rho_A = {\rm Tr}_{\overline{A}} [\rho_{A\overline{A}}]
$).
The entropy $s_A$ is zero if and only if the state of A is pure, \emph{i.e.} $\rho_A = |\phi_A\rangle \langle\phi_A|$. At $T=0$, the state of the whole system is pure, $\rho_{A\overline{A}} = |\psi_{A\overline{A}}\rangle \langle\psi_{A\overline{A}}|$, and $s_A=0$ if and only if the global state factors as $|\psi_{A\overline{A}}\rangle = |\phi_A\rangle \otimes |\eta_{\overline{A}}\rangle$. Thus, a non-zero value of local entropy $s_A$ signals the presence of entanglement between $A$ and $\overline{A}$, and moreover it is a quantitative measure of that entanglement.  

At finite temperature, the entanglement entropy acquires thermal contributions and is contaminated by thermal entropy $s$~\cite{Cardy:2017}. It no longer measures quantum correlations only. Nevertheless entanglement can persist up to high temperature~\cite{Vedral:T2004, Anders:2007, amicoRMP2008}. 

In the study of phases of correlated {\it fermionic} systems, the local, \emph{i.e.} onsite, entropy emerged as a powerful tool to identify phase transitions~\cite{zanardi2002, Gu:2004, Anfossi_Giorda_Montorsi_Traversa_2005, Anfossi:2007, larssonPRL:2005, larssonPRA2006, amicoRMP2008, byczukPRL2012, lanataCe, lanataEnt}. Therefore we focus on such a measure. First we discuss how we calculate this quantity, then we discuss its behavior in the supercritical region beyond the Mott endpoint.

\begin{figure*}
\centering{
\includegraphics[width=1\linewidth]{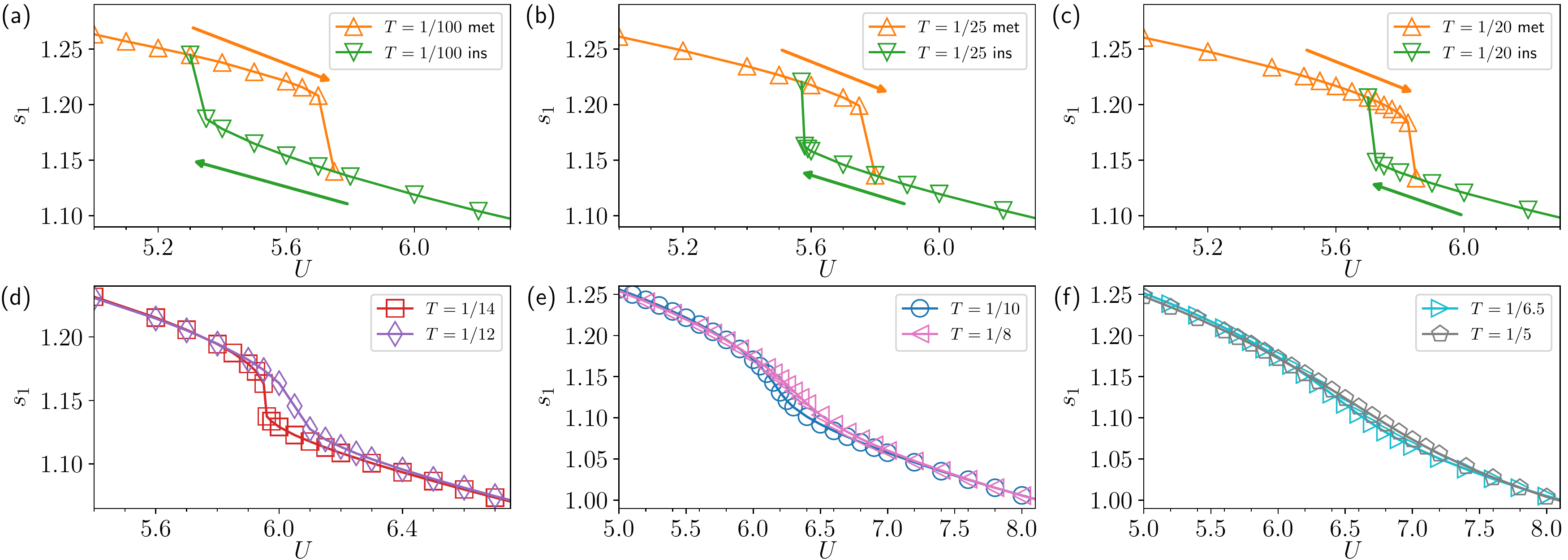}
}
\caption{Local entropy $s_1$ versus $U$ for different temperatures. Symbols are the same as in Fig.~\ref{figS-docc}. 
}
\label{figS-localentropy}
\end{figure*}

\subsubsection{Constructing local entropy}
Let $A$ be a site of the lattice and $B$ the remaining sites. The state-space of a single site is spanned by $\{ \ket{0}, \ket{\uparrow}, \ket{\downarrow}, \ket{\uparrow\downarrow} \}$. Because of particle-number and angular-momentum conservation, the reduced density matrix is then diagonal~\cite{zanardi2002}, 
\begin{align}
    \rho=p_0\ket{0}\langle 0| +p_\uparrow \ket{\uparrow}\langle \uparrow\!|+p_\downarrow  \ket{\downarrow}\langle \downarrow\!| +p_{\uparrow\downarrow} \ket{\uparrow\downarrow}\langle \uparrow \downarrow\!|, 
\end{align}
where $p_i$, with  $i=\{ 0, \uparrow, \downarrow, \uparrow\downarrow \}$, is the probability for a site to be empty, occupied with a spin up or down particle or doubly occupied. 
One finds 
\begin{align}
    p_{\uparrow \downarrow}&=\langle n_{i\uparrow} n_{i\downarrow} \rangle=D\\
    p_{ \uparrow} &= p_{\downarrow} = \langle n_{i\uparrow} -n_{i\uparrow} n_{i\downarrow} \rangle\\
    p_{0}&=1-2p_\uparrow -p_{\uparrow\downarrow}.
\end{align}
Thus, $s_1$ takes the form 
\begin{align}
s_1 &= -\sum_i p_i \ln(p_i) \\
 =& - (1-n+D) \ln(1-n+D) \\
 &-2\left(\frac{n}{2}-D \right) \ln\left(\frac{n}{2}-D\right) -D\ln(D).
\end{align}
In particular, at half-filling (i.e. $n=1$), we have,
\begin{align}
s_1 &= -2D\ln(D) -(1-2D) \ln(\frac{1}{2}-D),  
\end{align}
\emph{i.e.} $s_1$ is a function of double occupancy only. Knowledge of $D$ allows us to calculate $s_1$ directly. 

Recently, Ref.~\onlinecite{Wang:2014} describes a new algorithm to calculate the entanglement entropy (and R\'enyi entropies) for interacting electrons directly within the CTQMC impurity solver. It would be interesting to generalize their method for our problem in future investigations.

\subsubsection{Supercritical crossover of the local entropy}
\label{subsec:s0cross}

Let us turn to our results. In our companion letter~\cite{walshSl} we study the relation between local entropy $s_1$ and Mott transition. The key result is that $s_1$ detects the Mott transition and its supercritical crossover: for $T<T_c$, $s_1(U)_T$ identifies the first-order character of the transition by hysteretic behavior; for $T=T_c$ it shows critical scaling, and for $T_c$ $s_1(U)_T$ identifies supercritical crossover by sharp variations with $U$ marked by an inflection point.   
In the companion letter, the behavior of $s_1$ as a function of $U$ is shown only for a restricted range of temperatures. However, we did not shown that, analytically, the loci of inflections in $D(U)_T$ and $s_1(U)_T$ do not coincide in general, although the are numerically extremely close. Here we complete the picture. 

Figure~\ref{figS-localentropy} shows $s_1(U)_T$ at different temperatures. Our companion letter~\cite{walshSl} indicates that the key feature at $T>T_c$ is an inflection point. 
Our numerical data indeed shows that $s_1(U)_T$ has an inflection point whose tangent becomes infinite on approaching $T_c$. We show that for $T>T_c$ the inflection point in $s_1(U)$ does {\it not} coincide with that in $D(U)$, whereas at $T_c$ a singularity develops in both functions at the same $U_c$. A necessary condition to have an inflection point is that ${d^2s_1}/{dU^2}=0$. Now, 
\begin{align}
\frac{ds_1}{dU} & = \frac{ds_1}{dD}\frac{dD}{dU}  \label{der},
\end{align}
and thus
\begin{align}
\frac{d^2s_1}{dU^2} & =  \frac{d^2s_1}{dD^2} \, \left( \frac{dD}{dU} \right)^2 +\frac{ds_1}{dD} \, \frac{d^2D}{dU^2}.
\end{align}
For $T>T_{c}$, where $D(U)$ has an inflection point, \emph{i.e.} ${d^2D}/{dU^2} =0$, we have that $dD/dU$ is finite. 
Since $\frac{d^2s_1}{dD^2}=\frac{2}{D(1-2D)}$ is always positive, ${d^2s_1}/{dU^2}\ne 0$ so that $s_1$ does not have an inflection point there. Since our numerical data show that $s_1(U)$ has an inflection point at some value of $U$, this means that the inflection point in $s_1(U)$ does not coincide with that in $D(U)$. We find numerically that the two inflection points are very close (see line with blue crosses, and line with open red circles in Fig.~\ref{figS-cfr}(g)). 

However, at $T=T_c$, $D(U)$ becomes singular,
\begin{align}
\lim_{U\rightarrow U_{c}^\pm} \frac{dD}{dU} = -\infty,
\end{align}
so that, using the chain rule Eq.~(\ref{der}), we have
\begin{align}
\lim_{U\rightarrow U_c^\pm} \frac{ds_1}{dU} = \lim_{U\rightarrow U_c^\pm} \frac{ds_1}{dD}\frac{dD}{dU} = \frac{ds_1}{dD}  \lim_{U\rightarrow U_c^\pm} \frac{dD}{dU} = -\infty . 
\end{align}
where we used that $ds_1/dD$ is finite and strictly positive (apart from $D=0$ and $D=1/4$, which are far from $(U_{\rm c}, T_{\rm c})$). Therefore, at $T_{\rm c}$, $s_1(U)$ is singular in all is derivatives and the singularity coincides with that of $D(U)$.

The close numerical proximity between the inflection points in $s_1(U)$ and in $D(U)$ for a large range of temperatures is, in a sense, surprising because the inflections in $s_1(U)$ and $D(U)$ are not mathematically trivially identical. From another point of view, this is not too surprising since we know that, although the extrema in different response functions do not need to overlap for all temperatures, they need to converge at the critical endpoint. 

Physically, the largest change in magnitude of $s_1(U)$ found in close proximity with the largest fluctuations of the double occupancy $-dD/dU$, is telling us that large fluctuations in double occupancy lead to large changes in occupation probabilities.  
Furthermore, our results show that small variations in the interaction strength $U$ produce sharp changes in the $s_1(U)$, suggesting the idea of controlling entanglement properties close to the Mott transition to create an entanglement switch~\cite{OsterlohNature2002, OsbornePRA2002}. 

Finally, we note that, similarly to double occupancy and to entropy, the entanglement entropy $s_1$ scales as $-{\rm sgn}(U-U_c) |\left(  U-U_{c} \right)|^{1/\delta}$, although in general, for example when there is a non-zero magnetization, $s_1(U)$ does not need to scale like $D(U)$\cite{larssonPRA2006}.

\subsection{Mutual information }

Next we turn to the concept of mutual information. In this subsection, we clarify the definition of (normalized) total mutual information given in our companion letter~\cite{walshSl}, explaining why we used it instead of the standard definition of mutual information. This discussion is intimately related to the deviations of local entropy from extensivity, allowing us to introduce a new sequence of entanglement entropies whose decay with the size of the entangled region would be related to deviations from extensivity and to mutual information. We also explain why we see conceptual differences between our approach to mutual information and that used in the experiment on ultracold atoms~\cite{Cocchi:PRX2017}, even though the final mathematical expressions are the same. We extend our understanding of the phase diagram as seen by quantum information by showing data that compares inflections in $s$, $s_1$, and $\overline I_1$ (Fig.~\ref{figS-cfr}). This contains the new information that the loci of inflections of $\overline I_1(U)$ numerically coincide with the loci of inflections in the thermal entropy $s(U)_T$, as shown in particular in Fig.~\ref{figS-cfr}(g). Finally, we also show that the local entropy of an isolated site, $s_1'$, which is of purely thermal origin with no quantum contribution, is very different from $s_1$, reinforcing the fact that entanglement contributes significantly to correlations, even at finite temperature.

\subsubsection{Definitions of mutual information }

Let us consider the definition of mutual information, $I(A:B)= s_A +s_B -s_{AB}$. Physically, the (quantum) mutual information contains all quantum and classical correlations between $A$ and $B$. 
Subextensivity of entropy guarantees that the mutual information is non-negative, and it is non-zero only when the sum of the entropy of the parts exceeds the entropy of the whole. Information-theoretically, this means that the information lacking about the entire system is less than the sum of the information lacking about its parts, so some of the lacking information about the part must be common -- the parts are correlated~\cite{CoverInformation, watrous2018}.

We will be concerned with finite temperature, where it has been shown rigorously that, for a Hamiltonian with finite-range hoppings and interactions, mutual information scales as the size of the boundary of the two regions~\cite{wolfPRL2008}.

In Ref.~\onlinecite{Cocchi:PRX2017}, it was suggested that the mutual information can be obtained as follows. Consider one site, so that $s_A=s_1$ and $s_{AB}=sN$. Then, $s_B$ was approximated by $s_B=(N-1)s$, leading to $I=s_1-s$, the same expression as the one we used. However, this neglects corrections of order $1$ to $s_B$ because the entropy $s$ is extensive up to corrections of order $1/N$. At zero temperature, the result expected from the Schmidt decomposition $s_B=s_A=s_1$ comes precisely from a correction of order $1$ to $s_B$. With the definition $I_{AB}= s_A +s_B -s_{AB}$, the $T=0$ result $s_B=s_A=s_1$ should be $I= 2 s_1$, whereas $I=s_1-s$ leads to $I=s_1$. 
This leads us to define the notion of total mutual information.

\subsubsection{$s_1-s$ as total mutual information}

Suppose we are interested in the total mutual information between a given site $i$ and the rest of the lattice. Following the standard definition of mutual information, the mutual information between site $i=1$ and the rest of the lattice is $I(1:\{>1\}) = s_1+s_{\{>1\}} - s_{\{>0\}}$ where, following the companion letter~\cite{walshSl}, we denote by $\{>k\}$ the set of sites with indices greater than $k$, so $\{>0\}$ is the entire lattice. If we now consider the site labeled $i=2$, the mutual information between $i=2$ and the rest of the lattice $\{1\}\cup \{>2\}$ would lead to double-counting the correlations between sites 1 and 2. This mutual information has already been accounted for in the quantity $I(1:\{>1\})$. To avoid such double-counting, we first trace over site 1, that has already been considered. This gives us a new density matrix for sites $i=2$ and up. We can now, with this new density matrix, compute the mutual information between site $2$ and the remaining sites $\{>2\}$. Continuing this process, each time tracing out the $i-1$ sites already considered when we want the mutual information for site $i$, we define the total mutual information (normalized with $1/N$) between a single site and the rest of the lattice as
\begin{equation*}
\overline I_1 =  \frac 1N \sum_{i=1}^N I(i:\{>i\}) = \frac 1N \sum_{i=1}^N (s_1(i) + s_{\{>i\}} - s_{\{>i-1\}}).
\end{equation*}
Writing down the first few terms
\begin{align}
    \overline I_1 =  &\frac 1N (s_1(1) + s_{\{>1\}} - s_{\{>0\}}\\
                             &+s_1(2) + s_{\{>2\}} - s_{\{>1\}}\\
                             &+s_1(3) + s_{\{>3\}} - s_{\{>2\}}\\
                             &+ ...\\
                             &+s_1(N) + 0 - s_{\{>N-1\}}),
\end{align}
we see that most terms cancel, leaving $\overline I_1 =  ( \sum_{i=1}^N s_1(i)/N - s)$ where $s=s_{\{>0\}}/N$ is the thermodynamic entropy per site. For a translationally invariant system, all $s_1(i)$ are equal, so the total mutual information further simplifies to the difference between the local entropy and the thermodynamic entropy $\overline I_1 = s_1-s$. This quantity measures the total mutual information between one site and regions of all other possible sizes, avoiding overcounting mutual information with sites already considered.

\begin{figure*}
\centering{
\includegraphics[width=1.\linewidth]{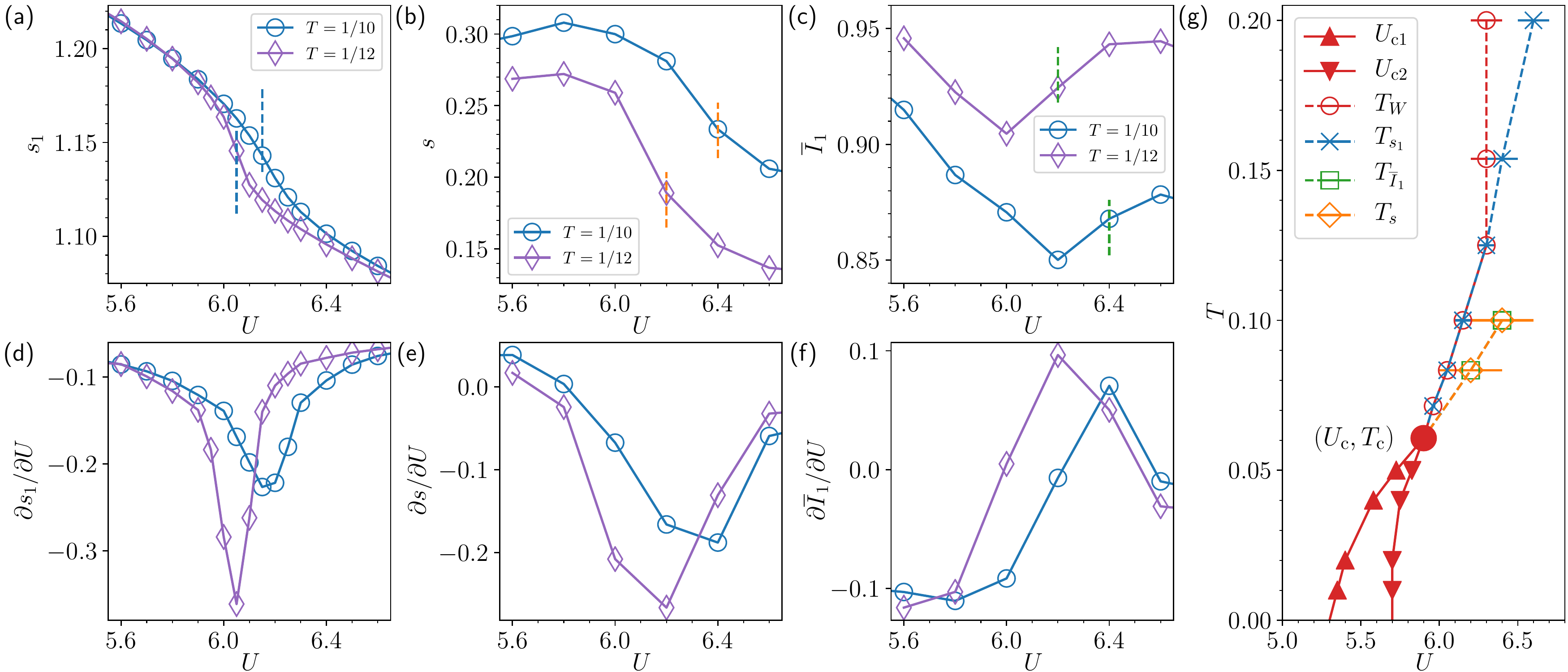}
}
\caption{(a) Local entropy $s_1$ versus $U$. (b) Thermodynamic entropy per site $s$ versus $U$. (c) Total mutual information $\overline I_1 = s_1 - s$ versus $U$. Vertical dashed lines in panels (a)-(c) mark the position of the inflection point. (d), (e), (f) Numerical derivative of $s_1$, $s$ and $\overline I_1$ with respect to $U$. Data in panels (a)-(f) are obtained for $T=1/10$ (blue circles) and $T=1/12$ (violet diamonds). (g) $T-U$ phase diagram of the 2D Hubbard model at $n=1$, including the loci of inflection points for the local entropy $s_1$ (blue crosses, $T_{s_1}$), for the total mutual information (green squares, $T_{\overline I_1}$) and for the thermodynamic entropy (orange diamonds, $T_{s}$). Other symbols as in Fig.~\ref{figS-PhaseDiagram}.}
\label{figS-cfr}
\end{figure*} 

\subsubsection{Supercritical crossover of the total mutual information }

Next, let us turn to our results. The companion letter~\cite{walshSl} shows that the Mott transition and its supercritical crossover are imprinted not only in the local entropy, but in the total mutual  information $\overline I_1$ as well. For $T<T_c$ $\overline I_1(U)_T$ detects the first-order nature of the transition by hysteretic behavior. At $T=T_c$ it reveals critical behavior. For $T>T_c$, $\overline I_1(U)_T$ shows non-monotonic behavior with minimum followed by a rapid increase marked by an inflection. The positions of this inflection for each temperature keeps track of the supercritical crossovers beyond the endpoint.

Figure~\ref{figS-cfr} shows data comparing the inflections in $s$, $s_1$ and $\overline I_1$. This adds to the companion letter the new result that the loci of inflections of $\overline I_1(U)$ numerically coincide with the loci of inflections in $s(U)_T$, as seen from green squares and orange diamonds in Fig.~\ref{figS-cfr}(g). Similarly to double occupancy and to entropy, the total mutual information $\overline I_1$ also scales as $\textrm{sgn}(U-U_c)|\left(  U-U_{c} \right)| ^{1/\delta}$. The difference between local entropy and thermodynamic entropy, $s_1-s$, quantifies correlations between a site and its environment. At $T=\infty$ where degrees of freedom become independent, $s=s_1$ and $\overline I_1=0$. At $T=0$, $s=0$ and $\overline I_1=s_1$. Further physical discussion is found in the companion letter~\cite{walshSl}.

\subsubsection{Local entropy with and without hybridization}

\begin{figure}[hb]
\centering{
\includegraphics[width=1.\linewidth]{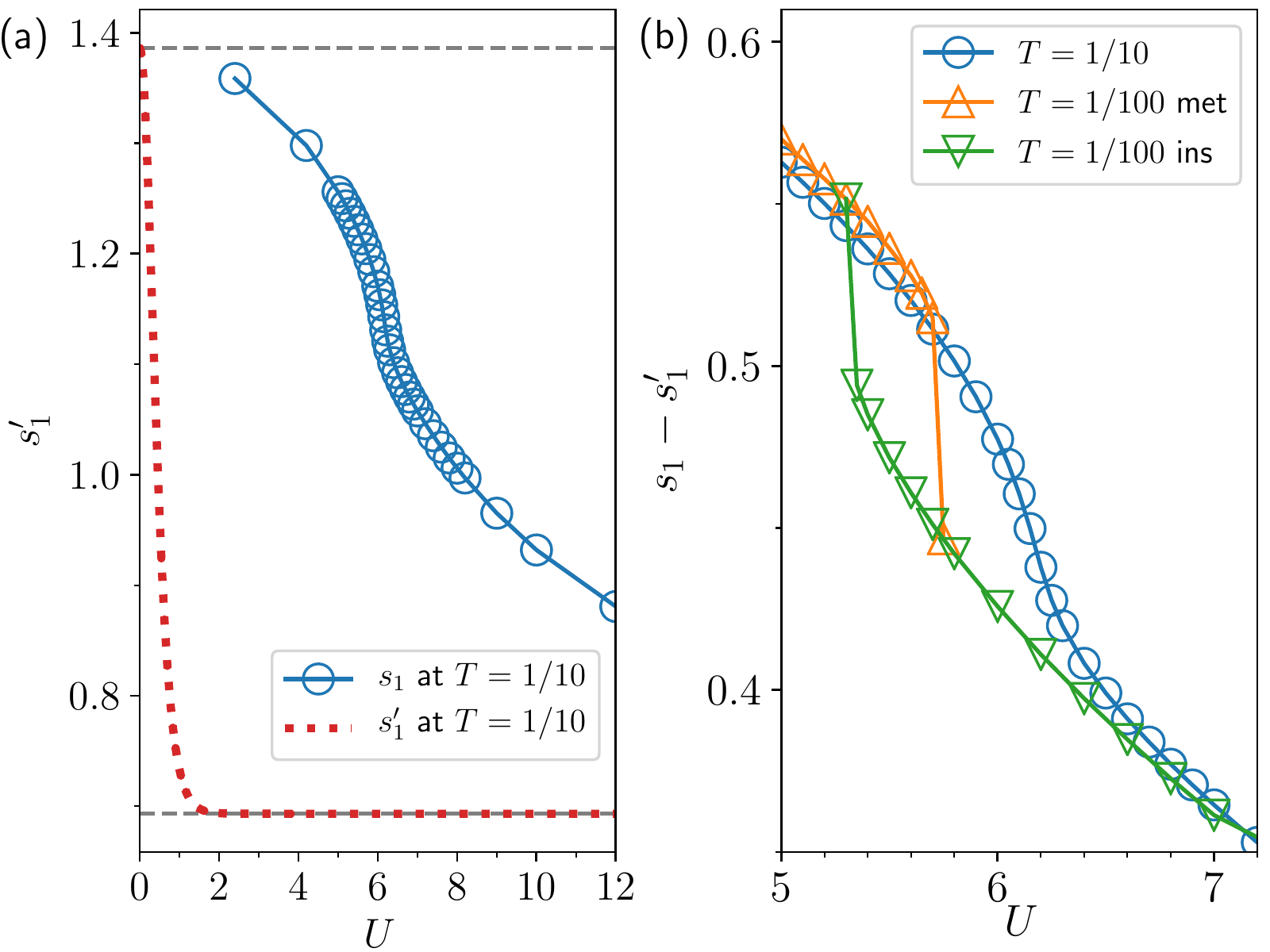}
}
\caption{(a) Local entropy $s_1$ versus $U$ at $n = 1$ and $T = 1/10$ (blue open circles). Horizontal lines mark $\ln4$ (for $U = 0$) and $\ln 2$ (for $U=\infty$). Dotted red line is the entropy of an isolated site in the grand canonical ensemble, $s_1'$. (b) Difference between local entropy and entropy of an isolated site in the grand canonical ensemble $s_1 - s_1'$ versus $U$ at $n=1$ for $T=1/10$ and $T=1/100$.}
\label{figS-sp}
\end{figure} 

Up to now, we have considered $s_1-s$, i.e. the difference between local entropy and entropy per site. From a complementary perspective, we can also compare the local entropy $s_1$ to the entropy of an isolated site (i.e. without hybridization to a bath) in the grand canonical ensemble, $s_1'$. The quantity $s_1'$ is for an isolated site with interaction $U$ in the grand-canonical ensemble. It contains only thermal contributions, no quantum contributions. The quantity $s_1-s_1'$ quantifies the contributions to the entropy from the terms coming from the hopping in and out of the bath, and thus from all the quantum and classical correlations between site and bath.  

Fig.~\ref{figS-sp}(a) shows $s_1'$ and $s_1$ versus $U$, whereas Fig.~\ref{figS-sp}(b) shows their difference. For $U > 0$, $s_1'(U)$ is smaller than $s_1(U)$, implying that the system with hybridization has less information than the isolated site. In other words, on average, the probability of having double occupancy is larger when the site is hybridized with the bath and the space of possible states on one site is larger than when there is no hybridization with a bath. In the latter case, double occupancy is quickly suppressed by $U$ and the information about the site is larger ($s_1'(U)$ smaller) since we know that the site is occupied on average by one electron and that other states are much less probable.

\subsubsection{Extension: deviations from extensivity and a sequence of entanglement entropies}

If we assume that at finite temperature, entanglement entropy has an extensive contribution plus a contribution proportional to the area, it is easy to understand why the standard definition of mutual information scales like the area of the common boundary~\cite{wolfPRL2008}. This hypothesis has another consequence. 

Consider the following sequence of entanglement entropies. Let $s_1$ be the entropy of a single site, $s_2$ the entropy of two sites, ..., $s_n$ the entropy of a ball of $n$ sites. The entropy per site $s$ is $s=\lim_{n \rightarrow \infty} \frac{s_n}{n}$. This can be rewritten as $\lim_{n \rightarrow \infty} \left(  \frac{s_n}{n} - s\right) =0$, which mathematically means that for every $\epsilon>0$ there exists an integer $N$ such that for $n\ge N$, $| \frac{s_n}{n} - s |  < \epsilon$. Physically, this means that the sequence $\frac{s_n}{n} - s$  ($\bar{I_1}$ being the first term of this sequence) converges to $0$ when we have reached a size $N$ sufficiently large to consider the subsystems of size $N$ as non-correlated ones. In other words, the sequence converges to zero as the ratio of area to volume if $N^{1/d}$ is larger than the correlation length of the problem. 

As a simple classical example, consider a chain of $N$ Ising spin $1/2$ dimers. Let us assume that in the dimers the spins are locked to point in the same direction. Among the different dimers, the spins can point in arbitrary directions. In other words, we have the mixed classical state $\rho=(\frac{1}{2}|\uparrow\uparrow\rangle \langle\uparrow\uparrow|+\frac{1}{2}|\downarrow\downarrow\rangle \langle\downarrow\downarrow|)^{\otimes N}$. Now, the entropy of a single spin, $s_1$, is $\ln 2$. The entropy of a dimer is $s_2=\ln 2$ as well. Therefore, for $n=2$, we found that $s_1 > s_2/2$, i.e. the entropy of a spin is larger than the entropy per spin. This is because of the correlations: within the dimer the spins are correlated. For $n\ge 2$, we also have $s_n/n=1/2$. Therefore for $n \ge 2$ even, there are no more correlations. In information language, the uncertainty per spin is smaller than the uncertainty on a single spin. 

Therefore $s_n/n-s$ measures the size-dependent non-extensivity of the entropy, and $s_1-s$ is a measure of the entropy due to correlations on the shortest distance. 
This is compatible with the findings of Ref.~\onlinecite{Anders:2006}, where entanglement at high temperatures can be detected by probing smaller and smaller parts of the system. It would be interesting to measure $s_n/n-s$ to see how the various correlation lengths (single-particle, spin, charge, etc.) control the size dependence of $s_n/n-s$.

\section{Conclusions}
\label{sec:Conclusions}

We revisited the iconic $T-U$ normal-state phase diagram of the half-filled 2D Hubbard model within plaquette CDMFT, with the goal of connecting thermodynamic concepts and information-theoretic ideas.  Other key motivations of our work are to advance our understanding of recent results with ultracold atoms in optical lattices~\cite{greinerNat2015, Cocchi:PRX2017} and to provide a path forward for new experiments. 

We improved the boundaries of the first-order Mott transition, and the location of the Widom-line in the supercritical region up to the percent accuracy level. We gave an exhaustive description of the thermodynamics near the Mott transition, revealing the behavior of pressure, charge compressibility, entropy, kinetic energy, potential energy and free energy across the Mott transition and its high-temperature crossovers. 

We found the so far unexplored first-order thermodynamic transition line $U_t(T)$ and showed that it is vertical and that it sits roughly in the middle of the previously identified spinodal lines when $T\rightarrow 0$. This allowed us to complete a study of the local thermodynamic stability of the coexisting metallic and insulating phases with a study of their global stability. We uncover binodal transition points and regions where either phase is unstable to nucleation of the other phase. Our analysis bears relevance for pioneering experiments addressing nucleations and metastability at the onset of the Mott transition~\cite{Ronchi:arXiv2018, Singer:2018, McLeod:2017, Qazilbash:2007, OCallahan:2016, Abreu:2015}. 

Calculation of the entropy from $n(\mu)$ using the Gibbs-Duhem relation is a methodological advance in the CDMFT context that can be exported to the study of other models. Physically, this calculation enabled us to show that the Widom line is also imprinted on the entropy and that the entropy has the expected critical scaling at the Mott endpoint. We found that the behavior of the entropy as a function of $U$ is highly non-monotonic, exhibiting a maximum, followed by an inflection point near the Widom line, followed by a minimum. This non-trivial behavior can be understood from the limiting $U\rightarrow 0$ and $U\rightarrow \infty$ behaviors and from the differences between the physics of elementary excitations in incipient metals and insulators. 

Knowledge of the entropy allowed us not only to obtain the grand potential and study global stability, it allowed us to compute the total mutual information between a single site and the rest of the system. The total mutual information $\overline I_1=s_1-s$ is a quantity that we introduced that is closely related to the usual concept of mutual information. Along with the local entropy $s_1$, we showed that it gives information-theoretic measures of correlations at the Mott transition. 
Here and in the companion letter~\cite{walshSl}, we uncovered their characteristic behaviors along the first-order phase boundary, near the critical endpoint, and in the supercritical region along the Widom line. Their sometimes unexpected behavior can be related to the physics of spin and charge excitations~\cite{walshSl}. We demonstrated that both $\overline I_1$ and $s_1$ can be used to detect the first-order Mott transition and the associated Widom line. This is a testable prediction for ultracold atom experiments~\cite{greinerNat2015, Cocchi:PRX2017}.
Finally, we suggested a sequence of entanglement entropies that are yet to be calculated, but that could provide new insights on the link between correlations and entanglement.

\appendix

\section{Critical behavior of the entropy at the Mott endpoint}
\label{sec:CriticalS}

\begin{figure}
\centering{
\includegraphics[width=1\linewidth]{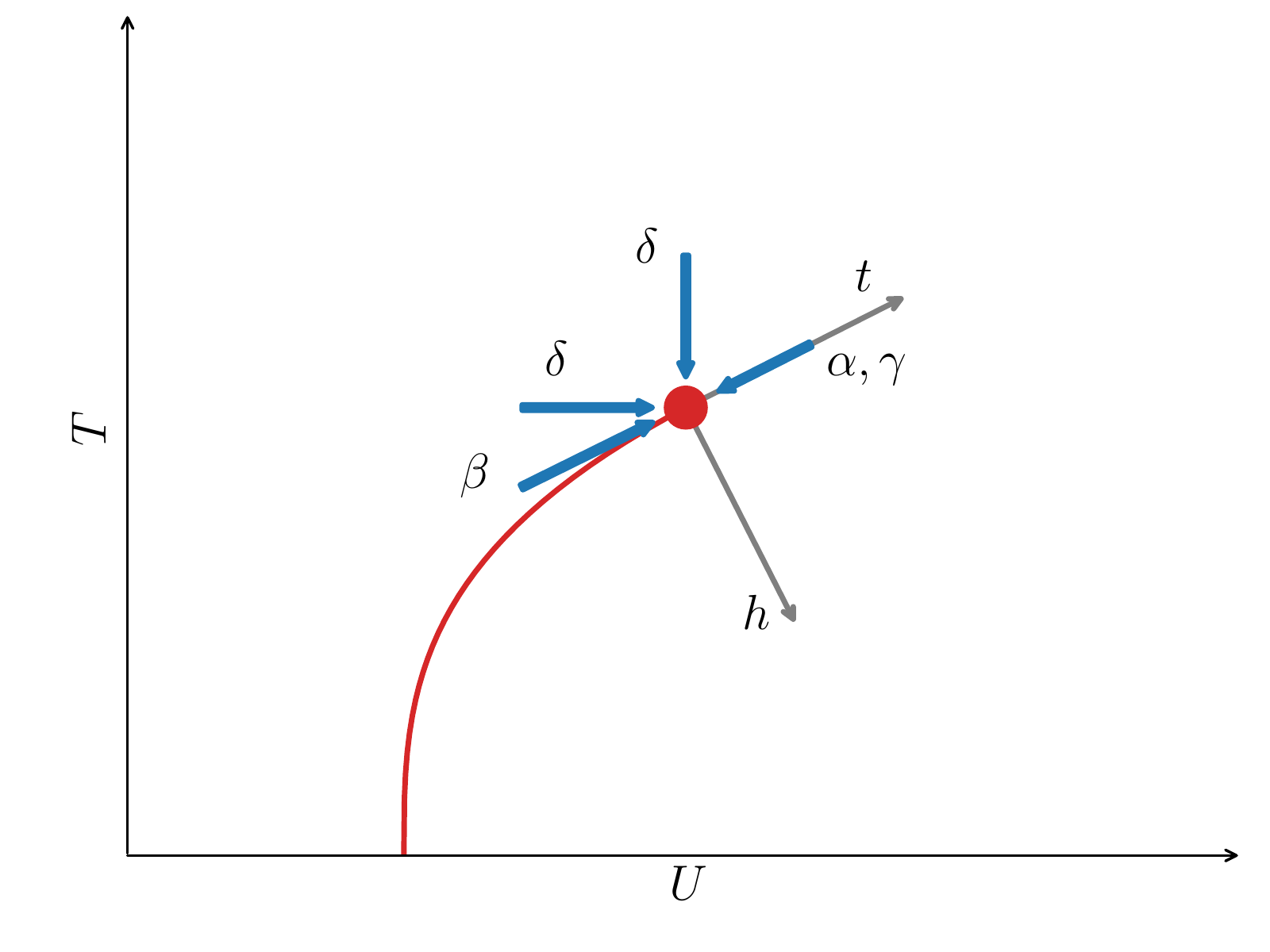}
}
\caption{Sketch of the interaction-driven Mott transition in the $T-U$ plane. The first-order transition (continuous red line) terminates in a second-order endpoint at $(U_c, T_c)$ (red circle). Gray arrows indicate the eigendirections $t$ and $h$. Blue arrows indicate the critical exponents along directions of interest here.
}
\label{figS-cr}
\end{figure}

The first-order Mott transition terminates in a critical endpoint. Close to that point, first derivatives and second derivatives of the singular part of the free energy $f$ vanish or diverge as power laws. The power is the critical exponent. Common critical exponents are $\alpha, \beta, \gamma, \delta$, which are universal and related via scaling laws. To calculate the exponents, one needs to indicate along which direction one is approaching the critical endpoint, i.e. if one is approaching the critical endpoint along the temperature-like and magnetic-field like renormalization-group eigendirections $t$ and $h$, or if one is approaching the endpoint from other directions. This is because it is not known a priori whether the eigendirections $t$ and $h$ are aligned with the physical coordinates temperature $T$ and interaction strength $U$. Figure~\ref{figS-cr} shows a sketch of the Mott transition with the eigendirections. 

In this section we consider the critical behavior of the entropy $s$. Ref.~\onlinecite{patrickCritical} shows that within CDMFT, double occupancy scales as $-{\rm sgn}(U-U_c)|U-U_c|^{1/\delta}$, with $\delta=3$.

\subsection{Scaling of the entropy along physical coordinates}

Les us begin with a mean-field point of view. The Ginzburg-Landau free energy functional for the order parameter $\eta$ takes
the form
\begin{equation}
f=t\eta^{2}+c\eta^{4}+h\eta, 
\end{equation}
where $t$ and $h$ are, respectively, the temperature-like and magnetic-field like eigendirections, and $c$ is a constant.
If these eigendirections are not aligned with the temperature and interaction axis, then
\begin{align}
t  & =t_{1}\left(  U-U_{c}\right)  +t_{2}\left(  T-T_{c}\right)  \nonumber\\
h  & =h_{1}\left(  U-U_{c}\right)  +h_{2}\left(  T-T_{c}\right)
\end{align}
and the entropy is given by
\begin{equation}
s=-\frac{\partial f}{\partial T}=-\left(  t_{2}\eta^{2}+c\eta^{4}+h_{2}%
\eta+\frac{\partial f}{\partial\eta}\frac{\partial\eta}{\partial T}\right)
.\label{Entropy}%
\end{equation}
In equilibrium, $\eta$ is given by $\frac{\partial f}{\partial\eta}=0$: 
\begin{equation}
2t\eta+4c\eta^{3}+h=0.
\end{equation}
Approaching the transition along the line $T=T_{c}$ and $\left(U-U_{c}\right)  $ the latter equation becomes 
\begin{equation}
2t_{1}\left(  U-U_{c}\right)  \eta+4c\eta^{3}+h_{1}\left(  U-U_{c}\right)  =0.
\end{equation}
To leading order,
\begin{equation}
\eta\approx-\frac{h_{1}}{4c}\left(  U-U_{c}\right)  ^{1/3}.
\end{equation}
Substituting in the equation for entropy, the leading order is
\begin{equation}
s\approx\frac{h_{2}h_{1}}{4c}\left(  U-U_{c}\right)  ^{1/3}.
\end{equation}
This is the mean-field behavior that is expected from dynamical mean-field theory~\cite{patrickCritical}. In general, we should have 
\begin{equation}
s\approx\frac{h_{2}h_{1}}{4c}\left(  U-U_{c}\right)  ^{1/\delta}.
\end{equation}

\subsection{Scaling of the entropy along an eigendirection}

The free energy obeys the following scaling relation: 
\begin{equation}
f\left(  \lambda^{p}t,\lambda^{q}h\right)  =\lambda^{d}f\left(  t,h\right).
\end{equation}

Consider the entropy $s=-\frac{\partial f\left(  t,h\right)  }{\partial t}$. Then
\begin{align}
\lambda^{p}\frac{\partial f\left(  \lambda^{p}t,\lambda^{q}h\right)
}{\partial\left(  t\lambda^{p}\right)  }  &  =\lambda^{d}\frac{\partial
	f\left(  t,h\right)  }{\partial t}\\
\lambda^{p}s\left(  0,\lambda^{q}h\right)   &  =\lambda^{d}s\left(
0,h\right)
\end{align}
and choosing $\lambda=h^{-1/q}$, we have
\begin{equation}
h^{\frac{d-p}{q}}s\left(  0,1\right)  =s\left(  0,h\right)  .
\end{equation}
All we need to do is rewrite this in terms of known exponents, that are
related to $p$ and $q$. Since we have the equalities
\begin{align}
p  &  =1/\nu\\
q  &  =\frac{1}{2}\left(  d+2-\eta\right),  
\end{align}
when we replace $h$ by $h\sim\left(  U-U_{c}\right)  ,$ the scaling of the entropy becomes
\begin{equation}
s\left(  0,\left(  U-U_{c}\right)  \right)  \sim\left(  U-U_{c}\right)
^{\frac{2\left(  d-\frac{1}{\nu}\right)  }{d+2-\eta}}.
\end{equation}
Since $\eta$ is not an exponent that is frequently used, we manipulate this to have an expression that involves better known exponents:
\begin{equation}
d+2-\eta=\delta\left(  d-2+\eta\right)  =\delta\frac{2\beta}{\nu}. 
\end{equation}
so that we can rewrite the result in the simpler form
\begin{equation}
s\sim\left(  U-U_{c}\right)  ^{\frac{\left(  d\nu-1\right)  }{\beta\delta}}.
\end{equation}
Since $\alpha=2-\nu d$, we find the final form
\begin{equation}
s\sim\left(  U-U_c\right)  ^\frac{1-\alpha}{\beta\delta}.
\end{equation}

For $d=2$, the Onsager solution gives
\begin{equation}
\frac{1-\alpha}{\beta\delta}=\frac{1-0}{\frac{1}{8}\ast15}=\frac{8}{15}
\end{equation}
while for $d=3$~\cite{Pelissetto:2002}, 
\begin{equation}
\frac{1-\alpha}{\beta\delta}=\frac{1-0.110}{0.3265\ast4.789} \approx 0.5692.
\end{equation}

Hence, the entropy is continuous at $U=U_{c}$ but its first derivative and all
others with respect to $U$ are singular.

\begin{acknowledgments}
We acknowledge Janet Anders and Marcelo Rozenberg for useful discussions. This work has been supported by the Natural Sciences and Engineering Research Council of Canada (NSERC) under grants RGPIN-2014-04584, RGPIN-2014-06630, the Canada First Research Excellence Fund and by the Research Chair in the Theory of Quantum Materials. Simulations were performed on computers provided by the Canadian Foundation for Innovation, the Minist\`ere de l'\'Education des Loisirs et du Sport (Qu\'ebec), Calcul Qu\'ebec, and Compute Canada.
\end{acknowledgments}


%

\end{document}